\documentclass[conference]{IEEEtran}
\IEEEoverridecommandlockouts
\usepackage{cite}
\usepackage{amsmath,amssymb,amsfonts}
\usepackage{algorithmic}
\usepackage{graphicx}
\usepackage{textcomp}
\usepackage{mathrsfs}
\usepackage{amsmath}
\usepackage{breqn}
\usepackage{color}
\usepackage{multirow}
\usepackage{xcolor}
\usepackage{subfigure}
\usepackage{colortbl}
\definecolor{mygray}{gray}{.9}
\usepackage{threeparttable}
\usepackage{color, soul}
\usepackage{ulem}
\usepackage{mathtools}
\usepackage{tikz}
\usepackage{textcomp}
\usepackage{lipsum}
\DeclarePairedDelimiter\ceil{\lceil}{\rceil}

\newcommand{\stc}[1]{{\color{black}#1}}

\usepackage{ulem}
\def\BibTeX{{\rm B\kern-.05em{\sc i\kern-.025em b}\kern-.08em
    T\kern-.1667em\lower.7ex\hbox{E}\kern-.125emX}}

\newcommand\copyrighttext{%
  \footnotesize \textcopyright 2024 IEEE. Personal use of this material is permitted.
  Permission from IEEE must be obtained for all other uses, in any current or future
  media, including reprinting/republishing this material for advertising or promotional
  purposes, creating new collective works, for resale or redistribution to servers or
  lists, or reuse of any copyrighted component of this work in other works.
  DOI: 10.1109/TVLSI.2023.3268084.}
  
\newcommand\copyrightnotice{%
\begin{tikzpicture}[remember picture,overlay]
\node[anchor=south,yshift=10pt] at (current page.south) {\fbox{\parbox{\dimexpr\textwidth-\fboxsep-\fboxrule\relax}{\copyrighttext}}};
\end{tikzpicture}%
}

\begin{document}

\title{
COAC: Cross-layer Optimization of Accelerator Configurability for Efficient CNN Processing}

\author{\IEEEauthorblockN{
Steven Colleman}
\IEEEauthorblockA{\textit{ESAT-MICAS, KULeuven} \\
Leuven, Belgium \\
steven.colleman@esat.kuleuven.be}
\and
\IEEEauthorblockN{
Man Shi}
\IEEEauthorblockA{\textit{ESAT-MICAS, KULeuven} \\
Leuven, Belgium \\
man.shi@esat.kuleuven.be}
\and
\IEEEauthorblockN{
Marian Verhelst}
\IEEEauthorblockA{\textit{ESAT-MICAS, KULeuven} \\
Leuven, Belgium \\
marian.verhelst@esat.kuleuven.be}
}

\maketitle
\copyrightnotice

\begin{abstract}
To achieve high accuracy, convolutional neural networks (CNNs) are increasingly growing in complexity and diversity in layer types and topologies. This makes it very challenging to efficiently deploy such networks on custom processor architectures for resource-scarce edge devices. Existing mapping exploration frameworks enable searching for the optimal execution schedules or hardware mappings of individual network layers, by optimizing each layer's spatial (dataflow parallelization) and temporal unrolling (execution order). However, these tools fail to take into account the overhead of supporting different unrolling schemes within a common hardware architecture. Using a fixed unrolling scheme across all layers is also not ideal, as this misses significant opportunities for energy and latency savings from optimizing the mapping of diverse layer types. A balanced approach assesses the right amount of mapping flexibility needed across target neural networks, while taking into account the overhead to support multiple unrollings. 
This paper, therefore, presents COAC, a cross-layer design space exploration and mapping framework to optimize the flexibility of neural processing architectures by balancing configurability overhead against resulting energy and latency savings for end-to-end inference. COAC does not only provide a systematical analysis of the architectural overhead in function of the supported spatial unrollings, but also builds an automated flow to find the best unrolling combination(s) for efficient end-to-end inference with limited hardware overhead. Results demonstrate that architectures with carefully optimized flexibility can achieve up to 38\% EDP (energy-delay-product) savings for a set of six neural networks at the expense of a relative area increase of 9.5\%.
\end{abstract}


\section{Introduction: literature survey and contribution}

\IEEEPARstart{M}{achine} learning algorithms nowadays carry more and more importance in a broad range of applications, such as image classification \cite{resnet} \cite{mobilenetv2} or acoustic processing \cite{park2016fully} \cite{yolov2}. 
The neural network layers constituting these networks vary widely in layer topologies: classical convolutional layers \cite{resnet}, pointwise and depthwise layers \cite{mobilenet}, attention layers \cite{attention}, etc. As illustrated in \cite{heterogeneous}, there are even large dissimilarities between layers of the same type, which can strongly differ in terms of channel, height, and width dimensionality. The channel-to-activation size ratio in that study varies between 0.002 and 4096 across different benchmarked networks. From a hardware point of view, this results in very diverging opportunities for spatial data reuse and parallelization in dedicated neural network processors (NPUs), denoted by the 'Spatial Unrolling' (SU) \cite{loma} of the processor's datapath. As a result, optimal hardware architectures vary depending on the targeted layer topologies. Many architectures, unfortunately, cannot exploit these optimal unrolling opportunities for all layers of interest, as they lack flexibility and only support a fixed SU for end-to-end inference, like the Edge TPU \cite{edge}. As illustrated in \cite{spatially}, using such a fixed unrolling results in a low utilization  of the datapath's processing elements (PEs) in case there is a wide variety of layers in the network under execution. \\
\begin{figure}[tb]
\centering
\includegraphics[width=0.9\linewidth]{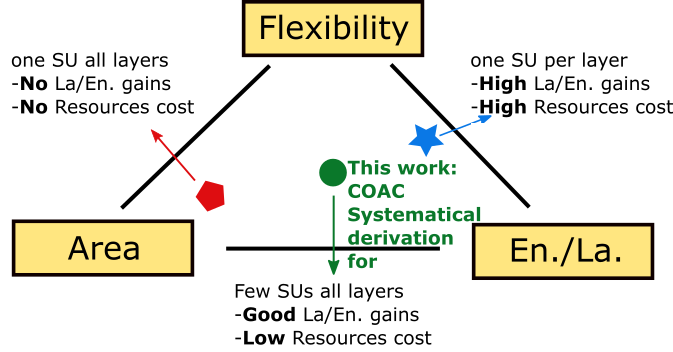}
\caption{Problem statement with our contribution: systematically derive combination of SUs to be able to exploit variability with minimizing resources overhead.} 
\label{fig:problem}
\end{figure}
Therefore, several studies, such as \cite{spatially} and~\cite{diana}, propose to deploy two different DNN accelerators  within the same processing system, each with a different SU to achieve higher PE utilization for diverse types of layers: e.g., one optimized for the depthwise layers and one optimized for the non-depthwise layers. These two PE arrays can work in parallel, and in a pipelined way. In such architectures, both the total PE utilization and the utilization per PE array are increased due to the heterogeneous accelerator combination, which convincingly showed the benefits of supporting multiple SUs. Nevertheless, the architecture's utilization is only good for networks that contain specific layer combinations that exploit both cores. E.g., the heterogeneous architecture proposed in \cite{spatially} excels in the presence of convolutional and depthwise layers, while suffering lower performance for networks without depthwise layers, such as ResNet \cite{resnet}, as for them, the accelerator optimized for depthwise layers sits idle.

Other works further extend this idea to provide increased flexibility into the accelerator's supported spatial unrollings: the authors of \cite{heterogeneous} propose to physically instantiate a large number ($> 2$) of PE arrays onto the same die, each of them with a different SU and according independent memory hierarchy. This allows to optimally allocate layers to the different PE arrays, where each PE array can perform computations for a different network (layer) in parallel. However, depending on the actual workloads to be executed, some PE arrays will remain idle. Due to this, heterogeneous multi-core accelerators are often not attractive for resource-limited edge devices.

To decrease area overhead, other designers suggest to use a single accelerator, but make it reconfigurable across multiple supported SUs \cite{eyeriss, highutilisation, Flagship, evolver, diana, us}.
However, it is challenging to design an efficient reconfigurable core, as additional hardware logic and data transformation overheads between layers have to be taken into account when multiple SUs are supported in one processor core. Recent state-of-the-art reconfigurable core architectures empirically limit the overhead, by constraining the reconfigurability of the accelerator core. A well-known example is Evolver \cite{evolver}, which supports a limited series of similar SUs that do not require any data reshuffling between different execution passes. This, however, leads to a sub-optimal energy and latency solution for layers whose best unrolling is not part of the supported small set of SUs. \cite{diana} integrated two totally different SUs into one PE array and added a reshuffling buffer to transfer data between the different modes. This reshuffling buffer between the PE array and feature map memory, together with many MUXes, enables data transformation of the data layout between layers to comply with the SU requirements of the next layer. The two distinct SUs supported in Diana, however, are still insufficient to efficiently cover the abundant layer diversity across neural networks. 

One of the main issues in understanding how much SU flexibility is desirable is that none of the works in the state-of-the-art has provided an in-depth analysis of the relationship between the overheads of combining multiple SUs and the corresponding scheduling-enabled latency/energy benefits stemming from increased SU flexibility in the workload mapper. It is hence not clear 
what configurability can achieve the best trade-off point between the advantages and overheads of supporting more than one SUs within a neural accelerator. Existing frameworks like ZigZag \cite{zigzag}, Timeloop \cite{timeloop}, Maestro \cite{Maestro} and Interstellar \cite{Interstellar} can search for the optimal SU for individual layers of a network under a given processing element (PE) array size. Yet, they mainly focus on deriving the optimal unrolling for individual network layers, while neglecting the overhead in terms of additional hardware and data reshuffling cost between layers executed under different SUs. 
This would naively result in a different optimal SU per layer, with very large overheads to support the configurability and data reshuffling across all these different SUs, likely canceling all the aspired benefits from the layer-optimized unrollings. \\

This paper, therefore, aims to enable the exploration of the best combination of SUs across all the layers of a targeted (set of) neural networks. This is pursued through the development of a framework, named COAC, which can analytically assess and optimize the cross-layer latency/energy benefits, as well as the hardware overhead for supporting different SU combinations. Optimized search strategies enable to search the massive space of SU combinations in the latency/energy/area (resources) overhead space for end-to-end inference and balance the area overhead from flexibility against the benefits stemming from energy- and latency-optimal mapping opportunities of the targeted layer types.

To the best of our knowledge, this paper is the first work to explore and optimize this trade-off between the flexibility and the corresponding overheads of devising multiple SUs for one PE array with such a systematic approach to 
find different Pareto-optimal solutions in the latency/energy/overhead space.
\begin{figure}[tb]
\centering
\includegraphics[width=1\linewidth]{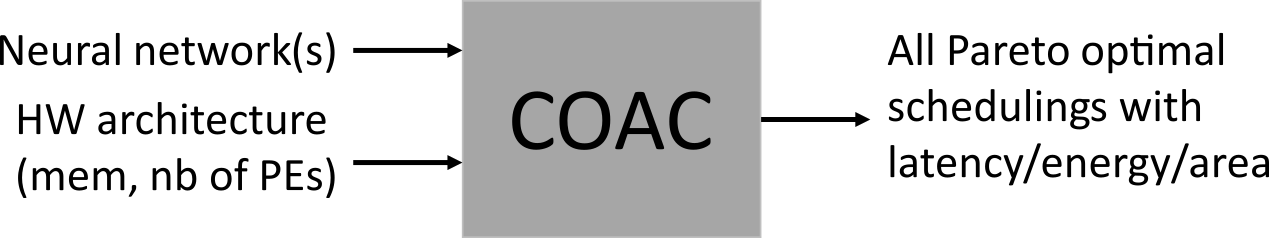}
\caption{Black box behavior of COAC. The internal working mechanism of COAC is explained throughout the paper.} 
\label{fig:intro}
\end{figure}
The presented COAC framework, as illustrated in Fig. \ref{fig:intro}, is compatible with state-of-the-art energy/latency estimation tools \cite{zigzag} \cite{timeloop}, as it adds a cross-layer analysis and optimization wrapper around the single layer evaluation frameworks. The experiments reported in this paper are conducted with ZigZag \cite{zigzag}, a SotA layer-wise design space exploration framework. The main contributions of this paper can be summarized as:
\begin{itemize}
    \item Introduction of a unified area overhead model that incorporates and quantifies the configurability cost of supporting different SUs in a common processing array
    (Section III). \stc{A template architecture containing three additional overhead blocks (data assignment block, output aggregation network and reshuffling buffer, as in \cite{evolver} \cite{diana}) in comparison with non-flexible hardware architectures is given.}
    The section also includes the analysis on how similarity between combined SUs can reduce this cost. 
    \item Realization of a systematic cross-layer design space exploration and mapping tool, COAC, to find the best combination of supported SUs for a given workload, taking flexibility overhead into account (Section IV). 
    \item Execution of case studies using COAC to show the optimal combination of SUs with their latency/energy/area results for a broad range of networks, and interpretation of the results to derive new general insights on optimal flexibility (Section V). 
\end{itemize}

\section{Background and motivation}
\begin{figure}[tb]
\centering
\includegraphics[width=1\linewidth]{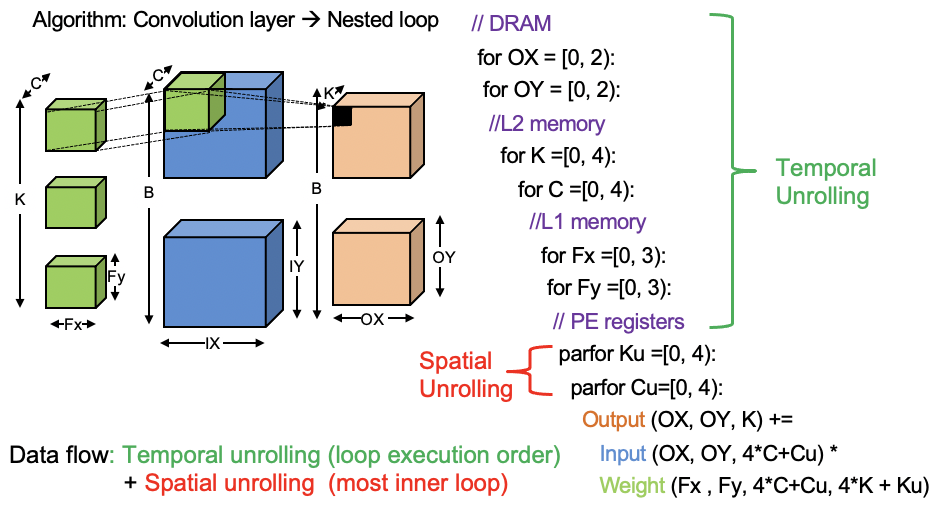}
\caption{Meaning of Spatial Unrolling versus Temporal Unrolling.} 
\label{fig:sutu}
\end{figure}

A CNN layer is defined by 5 or 6 nested for-loops ($Fx$ and $Fy$ for filter kernel, $Ox$ and $Oy$ for output feature dimensions, $C$ and $K$ for input and output channels. The latter are replaced by $G$ for groups in depthwise layers). 
After loop splitting and loop reordering, these loops can each be executed (unrolled) temporally (TU = temporal unrolling) or spatially (SU), as illustrated in Fig. \ref{fig:sutu}. This paper mainly focuses on the optimizing flexibility required for spatial unrolling. Different temporal unrollings can typically be supported more easily through software or FSM control.

\subsection{Spatial Unrolling on PE array} \label{begin}
\begin{figure}[tb]
\centering
\includegraphics[width=1.0\linewidth]{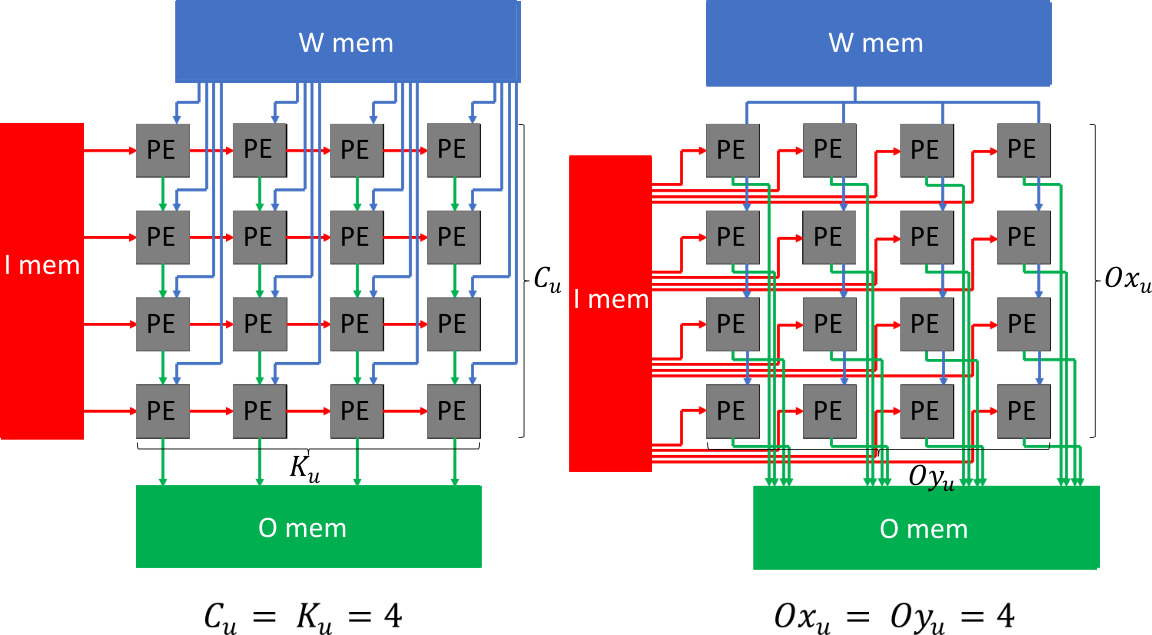}
\caption{Data distribution is completely different for different SUs.} 
\label{fig:illustrationimportance}
\end{figure}
Spatial Unrolling (SU) denotes the hardware parallelization over a loop parameter, indicating compute operations executing within the same clock cycle on a PE array. 
It is already important to realize that the support of specific SU's depends on the available data reuse and data distribution schemes within the PE array. These connectivity patterns determine the maximal spatial unrolling supported by the PE array along each loop parameter for layer mapping. This paper will use the suffix 'u' to indicate this maximal spatial unrolling capabilities.
Fig \ref{fig:illustrationimportance} illustrates this for 2 distinct spatial arrays: The array on the left, allows a spatial unrolling of $Cu|Ku$, which can reuse the inputs along the different columns and accumulate outputs along different rows. The array on the right, is more suited for an $OXu|OYu$ unrolling, which can reuse the same single weight across the complete array, but requires separate input and output activation fetching/storing.

In the remainder of this paper, we define a supported maximal spatial unrolling by how it distributes the total number of available PEs ($nb_{PEs}$) over the different possible unrolling dimensions:
\begin{equation}
\label{eq1}
nb_{PEs} = Ox_u\cdot Oy_u\cdot Fx_u\cdot Fy_u \cdot G_u \cdot C_u \cdot K_u 
\end{equation}
E.g. in Fig \ref{fig:illustrationimportance}, $nb_{PEs} = 16$.

\subsection{SU combining benefits for varying workloads}
\begin{table}[!t] 
\centering
\caption{Impact of SU on PE utilization.}
\begin{threeparttable}
\begin{tabular}{|c||c|c|}
   \hline
   SU & TPU-like \cite{tpu}\tnote{1}  &  G_unrolling-like \tnote{2} \\
   \hline \hline
Layer\#29 ResNet101 \tnote{3}& 96.2\% & 0.7\%  \\
\hline
Layer\#2  MobileNetv2 \tnote{4}& 0.7\% & 100\% \\
   \hline
\end{tabular}
\label{table:firstresults}
 \begin{tablenotes}
        \scriptsize
        \item[1] TPU-like : [$C_u$ = 12, $K_u$ = 12]
        \item[2] G_unrolling-like : [$Fx_u$ = 3, $Fy_u$ = 3, $G_u$ = 16]
        \item[3] layer info. : [K = 384, C = 256, Ox = Oy = 13, Fx = Fy = 3]
        \item[4] layer info. : [G = 32, Ox = Oy = 112, Fx = Fy = 3]
      \end{tablenotes}
    \end{threeparttable}
\vspace*{-0.5cm}
\end{table}

It is clear that not every layer can fully utilize the PE array under a specific SU. E.g., when a layer with an input channel dimension $C$ would be spatially unrolled with a supported input channel unrolling of $Cu$, then that dimension will have an effective utilization of $$C_u \ceil*{\frac{C}{C_u}}$$. The PE spatial utilization ($S_{ut}$) for a given workload and a given SU (as defined in (\ref{eq1})) is hence:
\begin{equation}
S_{ut} = \frac{Ox \cdot Oy \cdot Fx \cdot Fy \cdot G}{ Ox_u \ceil*{\frac{Ox}{Ox_u}} Oy_u \ceil*{\frac{Oy}{Oy_u}} Fx_u \ceil*{\frac{Fx}{Fx_u}} Fy_u \ceil*{\frac{Fy}{Fy_u}} G_u \ceil*{\frac{G}{G_u}}}
\label{eq3}
\end{equation}
for a depthwise (DW) layer, and similar for a classical layer. 
The result is that, due to the diversities in topologies of CNN layers, not every layer can be executed efficiently under a common SU.
From the small experiment of Table \ref{table:firstresults}, it can be seen that different layer topologies strongly impact the PE utilization for a given SU, hence clearly resulting in different optimal SU's for each layer. Based on this, it is clear that combining SUs can bring significant performance improvements. 
This means that the architecture supports 2 or more different $Ox_u, Oy_u, Fx_u, Fy_u, G_u, C_u, K_u$ combinations, for example, an architecture that supports both illustrated SUs from Fig. 3. 
However, this comes at the cost of some overheads in terms of hardware resources as correct data transfer/orientation for each SU must propagate through the architecture. This will be discussed in Section III. 

\subsection{Impact of temporal unrolling on hardware utilization}
The total utilization of a PE array is not only defined by the SU. As illustrated in \cite{us}, also the temporal unrolling (TU) has impact on the total utilization (and therefore throughput and latency). This stems from the difference in the number of weights, inputs, and outputs that are accessed per clock cycle.
Assuming $p$ bits in each input and weight data word and $2p$ bits for intermediate outputs \cite{us}, the required number of data fetches per clock cycle when only considering the SU, amounts:
\begin{equation}
W_{needed,SU} = p \cdot C_u \cdot K_u \cdot Fx_u \cdot Fy_u 
\end{equation}
\begin{equation}
I_{needed,SU} = p \cdot C_u \cdot (Ox_u + Fx_u - 1) \cdot (Oy_u + Fy_u - 1)
\end{equation}
\begin{equation}
O_{needed,SU} = 2p \cdot K_u \cdot Ox_u \cdot Oy_u 
\end{equation}
However, depending on the Temporal Unrolling, not all these data items need to be refreshed every cycle. The innermost temporal for-loop influences which data elements can remain stationary, and hence do not consume any memory bandwidth.  Assuming a maximal memory bandwidth/port width tolerated for the weights ($PW_W$), input ($PW_I$) and output ($PW_O$) memories, this allows to compute the memory induced computational stalls, and hence the resulting temporal utilization ($T_{ut}$) (index next to the $"ut"$ subscript indicating the innermost for-loop) \cite{us}:
\begin{equation}
T_{ut,C} \approx min \left( 1, \frac{PW_W}{W_{needed,SU}}, \frac{PW_I}{I_{needed,SU}} \right)
\end{equation}
\begin{equation}
T_{ut,K} \approx min \left( 1, \frac{PW_W}{W_{needed,SU}}, \frac{PW_O}{O_{needed,SU}} \right)
\end{equation}
\begin{equation}
T_{ut,Ox/Oy} \approx min \left( 1,  \frac{PW_I}{I_{needed,SU}}, \frac{PW_O}{O_{needed,SU}} \right)
\end{equation}
\begin{equation}
T_{ut,G} \approx min \left( 1, \frac{PW_W}{W_{needed,SU}}, \frac{PW_I}{I_{needed,SU}}, \frac{PW_O}{O_{needed,SU}} \right)
\end{equation}
The spatial and temporal utilization have to be multiplied to obtain the total utilization. From this, it is clear that the TU also has a large impact on latency, as well as on the system energy consumption. 
As such, it is important to perform the assessment of the latency or energy performance of a given SU across all possible compatible TUs. As these different TUs can typically be supported in software and require minimal hardware support, they will not impact the hardware resource study of Section III. 

\section{Estimating SU reconfigurability overhead} 
Table \ref{table:abbr} contains a summary of all abbreviations and symbols used in this section. 
\begin{table}[!t]
\renewcommand{\arraystretch}{0.9}
\caption{Abbreviations and their meaning}
\vspace*{-0.3cm}
\label{table:abbr}
\centering
\begin{tabular}{|c|c|}
\hline
 \bf{Symbol} & \bf{Meaning} \\ \hline \hline
$C$ & number of input channels \\ \hline
$K$ & number of output channels \\ \hline
$G$ & number of channels in depthwise layer \\ \hline
$Ox/Oy$ & spatial output feature dimensions \\ \hline
$Ix/Iy$ & spatial input feature dimensions \\ \hline
$Fx/Fy$ & kernel filter dimensions \\ \hline
$SU$ & spatial unrolling \\ \hline
$TU$ & temporal unrolling \\ \hline
$MUX$ & multiplexer \\ \hline
$DW$ & depthwise \\ \hline
$O_{sum}$ & \begin{tabular}{c} number of data items to be \\  accumulated for one output feature \end{tabular} \\ \hline
$W_u$ & number of weights needed in parallel \\ \hline
$A_u$ & number of activations needed in parallel \\ \hline
$W_r$ & max value of $W_u$ over all SUs under study \\ \hline
$A_r$ & max value of $A_u$ over all SUs under study \\ \hline
$PW_i$ & physical port width of memory i \\ \hline
$nb_{PEs}$ & number of PEs in the PE array \\ \hline
$N_{adders}$ & number of 2-input adders in adder tree \\ \hline
$O_{MUX}$ & multiplexers at output aggregation network \\ \hline
$R_{i,j}^{cl}$ & data chunk size for reshuffling between SU i and j \\ \hline
$R_{i,j}^{min}$ & minimal value of $R_{i,j}^{cl}$ over all SU combinations \\ \hline
$REG_{buffer}$ & number of words in the reshuffling buffer \\ \hline
$MUX_{buffer}$ & \begin{tabular}{c} number of multiplexers at \\ the output of the reshuffling buffer \end{tabular} \\ \hline
\end{tabular}
\vspace*{-0.3cm}
\end{table}

\subsection{Introduction}
\begin{figure}[tb]
\centering
\includegraphics[width=0.7\linewidth]{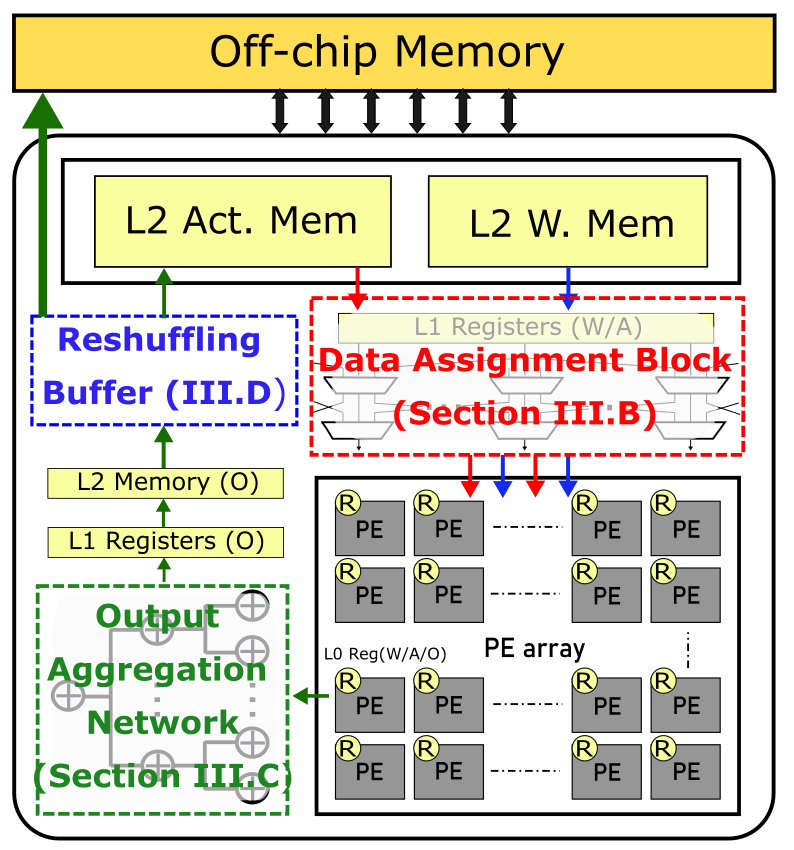}
\caption{\stc{Flexible Spatial unrolling (SU) template architecture, including the three overhead components (red/green/blue blocks): The data assignment block contains registers and MUXes, the output aggregation network contains MUXes and adders, the reshuffling buffer contains registers and MUXes. A more detailed illustration on how to configure these blocks can be found in Fig. \ref{fig:input3}.}} 
\label{fig:main}
\end{figure}
\begin{figure*}[tb]
\centering
\includegraphics[width=17cm, height=6.8cm]{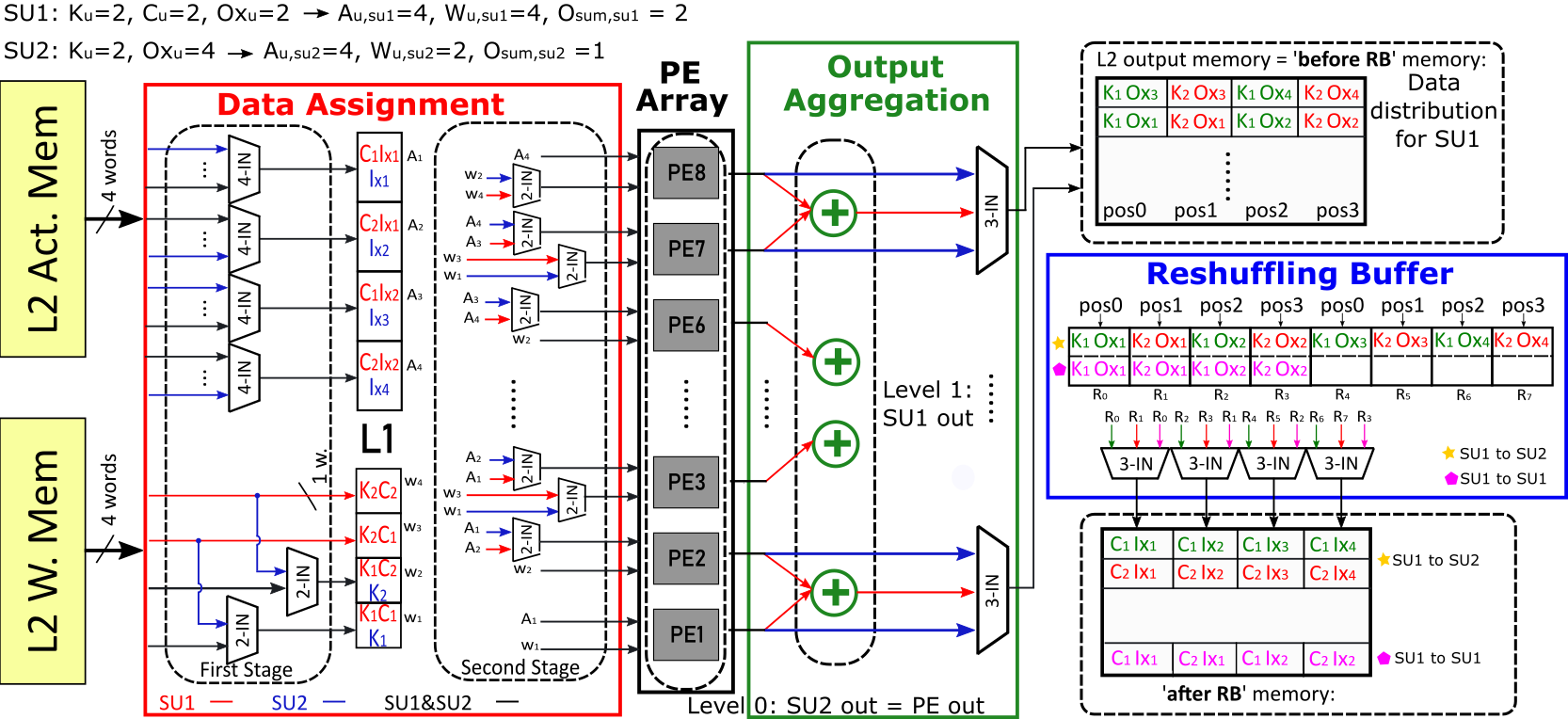}
\caption{Example for data assignment, output aggregation and reshuffling buffer. 
}
\label{fig:input3}
\end{figure*}
Fig. \ref{fig:main} depicts a hardware architecture template of a typical machine learning accelerator, extended with three required blocks necessary to support SU configurability: 
\begin{enumerate}
    \item a 
\textit{data assignment block} at the inputs of the PE array, consisting of multiplexers (MUXes, a $Q$ input MUX will be modelled as '$Q$ one-input MUXes' because we noticed that the area of a MUX increases linearly with the number of inputs) and registers (level L1) to ensure correct and reconfigurable input data routing (data shifting for e.g. padding); 
\item a \textit{reconfigurable adder tree} for flexible PE output aggregation, 
to support aggregation over different numbers of output features; and 
\item a \textit{reshuffling buffer} higher up in the output memory hierarchy to perform data reordering between subsequent layers with different SUs, to collocate activation data which will be consumed together and hence maximize data reuse and minimize memory stalls.
\end{enumerate}

In subsequent sections, we will carefully analyze the complexity of each of these three blocks in light of the supported combination of SUs. 
Define for a specific SU $O_{sum}$ as the number of data items to be accumulated within one clock cycle for one output feature:
\begin{equation}
O_{sum} = C_u \cdot Fx_u \cdot Fy_u
\end{equation}
Define $W_u$ for a specific SU as the number of weights needed in parallel:
\begin{equation}
W_u = G_u \cdot C_u \cdot K_u \cdot Fx_u \cdot Fy_u
\label{wu}
\end{equation}
This number is equal to the product of the SU factors for the loops that require unique weights for each part of the loop. 
Further define $A_u$ for a specific SU as the number of activations needed in parallel:
\begin{equation}
A_u = G_u \cdot C_u \cdot Ox_u \cdot Fx_u \cdot Oy_u \cdot Fy_u
\label{au}
\end{equation}
As done in many SotA architectures, we will assume data duplication at the L1 level is used when $Fx_u$ and $Fy_u$ are different from 1, and therefore different $(Ox, Oy)$ positions under computation require data from the same $(Ix, Iy)$ input position. 
This increases the number of L1 registers needed, but drastically simplifies design as will become clear later on. Furthermore, the template architecture assumes that input features of a given layer are stored in the L2 activation memory in groups of $Oy_u|Ox_u|C_u$ (resp. $Oy_u|Ox_u|G_u$ for a DW layer). It is the task of the reshuffling buffer (as discussed later) to group the data in this format, in case the previous' layer output did not comply with this.

Despite the wide variety of potential PE array implementations, a lower bound of configurability cost in function of PE array flexibility can be 
defined by making three reasonable assumptions:
\begin{itemize}
    \item The number of PEs is a power of 2. From Equation \ref{eq1}, it follows immediately that all unrolling factors will be powers of 2. \stc{This assumption is made for complexity reasons, and is in line with most SotA inference accelerators.}
    \item All data words, intermediate or final, are represented by a number of bits that is a power of 2. Additionally, the width of all memory ports is a power of 2 words.
    \item \stc{The architecture performs layer-by-layer CNN execution.}
\end{itemize}

Given these assumptions, the COAC framework can find the optimal SU combination for end-to-end inference of a given (suite of) workloads, taking into account the overhead for supporting the required SU configurability. The cost models for each of the configurability blocks depicted in Fig. \ref{fig:main} will first be explained in more depth in next subsections, using the example from Fig. \ref{fig:input3}, with 2 supported SUs (SU1 \& SU2, as defined in Fig. \ref{fig:input3}(top, left)) and all memories assumed to have a port width of 4 data words.

\subsection{Data assignment block cost analysis}
The data assignment block (Fig. \ref{fig:main}\&\ref{fig:input3}) first of all consists of a set of L1 registers that contains all W and A input words needed in parallel by the PE array, being $Oy_u|Ox_u|C_u$ resp. $Oy_u|Ox_u|G_u$ input data elements. These registers are preceded by a set of MUXes (first stage) to route each word of the L2 memory port to the correct L1 register. A second set of MUXes after the L1 registers ensures registered data words are routed to the right PEs for each supported SU. 

When multiple SUs are combined, the necessary number of L1 registers is determined by the maximal number of parallel input words required by the PE array across all supported SUs: $A_r = max(A_u)$ words for activations and $W_r =max(W_u)$ words for weights. In the example of Fig. \ref{fig:input3} two SUs (SU1 and SU2) are supported. Based on the equation of $Au$ and $Wu$ the required number of corresponding L1 activation and weight registers are $A_{u,SU1} = 4$, $A_{u,SU2} =4$ and $W_{u,SU1}=4$, $W_{u,SU2}=2$, respectively. Therefore, in this (simplified) example, $A_r = max(A_{u,SU1},A_{u,SU2}) = 4$ and $W_r = max(W_{u,SU1},W_{u,SU2}) = 4$.

Now, we will analyze the number of MUXes in the data assignment block. Define $z(x)$ as a function that returns 0 if $x = 1$ and $x$ itself otherwise.  
In general terms, the total number of one-input MUXes in the first stage for the weights assignment $W_{MUX,1}$ is
\begin{equation}
W_{MUX,1} = \displaystyle\sum_{i=1}^{W_r} z\left( \ceil*{ \frac{PW_{L2,weights}}{min(W_u), \forall \ SU_j | (W_{u,SU_j} \ge i)}} \right)
\label{wmux1}
\end{equation}
The total number of one-input MUXes in the first stage for the activations assignment $A_{MUX,1}$ is
\begin{equation}
A_{MUX,1} = \displaystyle\sum_{i=1}^{A_r} z\left( \ceil*{\frac{PW_{L2,act}}{ min(G_u \cdot C_u), \forall \ SU_j | (A_{u,SU_j} \ge i)}} \right)
\label{imux1}
\end{equation}
\stc{The derivation of both equations can be found in Appendix 1. This appendix also describes the algorithm to compute the number of MUXes in the second stage. Briefly summarized, the algorithm computes for each PE and for each SU which L1 register feeds data to the specific PE, and in this way counts the number of sources the data for a given PE can come from. }

\subsection{Output aggregation network cost analysis}

The data aggregation network at the output of the PE array accumulates partial results across $O_{sum}$ PEs for the different supported SUs (see Fig.~\ref{fig:main}\&\ref{fig:input3}). When supported SUs have different $O_{sum}$ values, the aggregation network will also need reconfigurability. As $O_{sum}$ is always a power of 2 and due to the fact that PEs belonging to the same $O_{sum}$ group are always neighboring PEs, the output aggregation network can be implemented as a multi-level reconfigurable adder tree. Depending on which SU is under execution, outputs from a different level of the hierarchical adder tree must be extracted. The depth of the adder tree 
is determined by the maximum value of $O_{sum}$ across all supported SUs, denoted as $O_{sum}^*$. The PE array will in total need $N_{adders}$ two-input adder units:\\
\begin{equation}
N_{adders} = (O_{sum}^* -1)  \cdot  \frac{nb_{PEs}}{O_{sum}^*}
\label{nadders}
\end{equation}
This is illustrated in the example of Fig. \ref{fig:adderuitleg}, where an output aggregation network is designed for a system with 8 PEs and 3 SUs. The values for $O_{sum}$ are respectively 1, 2 and 4. Therefore, $O_{sum}^* = 4$.

In general, the total number of two-input MUXes ($O_{MUX}$) in the output aggregation network can be formulated as:
\begin{equation}
O_{MUX} = PW_{L2,O} \cdot z\left( \displaystyle\sum_{i} f(i) \cdot max \left(\frac{\frac{nb_{PEs}}{2^i}}{PW_{L2,O}}, 1 \right) \right)
\label{eqadder}
\end{equation}
where $f(i)$ is 1 if level $i$ contains final outputs for at least 1 of the supported SUs and 0 otherwise. \stc{This equation is derived in Appendix 2.}

\subsection{Reshuffling buffer overhead modeling}
Once the output data arrives in this way in the L2 output memory (Fig.~\ref{fig:main}), it is not necessarily ready yet for consumption by the next layer, which can potentially make use of a different SU. Hence, additional overheads in terms of memory stalls and memory accesses energy have to be taken into account when different SUs are adopted for subsequent layers. In the example of Fig.~\ref{fig:input3}, assume the first layer is conducted based on SU1, and the result of this layer will be used for the next layer computed with SU2. For SU1, 4 outputs words (= $K_u \cdot Ox_u \cdot Oy_u$; $OX_u =2$, $K_u =2$) are computed in parallel and stored at the same memory line. SU2 needs 4 input words (= $C_u\cdot Ox_u \cdot Oy_u$; $OX_u =4$)  in parallel, which ideally also should be stored in the same memory line to avoid stalling cycles for data fetching. 
However, only 2 out of the 4 parallel computed outputs of SU1 (2 consecutive OXs from same K) belong to the same input group of SU2 (4 consecutive OXs of the same K). 
It is clear that not all data that are generated in parallel by SU1 are needed in parallel in SU2, and not all data that SU2 needs in parallel are computed and stored together in SU1. This will result in additional energy and latency penalties for data reorganization, that may kill the advantages of supporting multiple SUs when not properly resolved. 

Stalling can be avoided by exploiting a reshuffling buffer within the memory hierarchy between the L2 output activation memory and the L2 input activation memory (Fig.~\ref{fig:main} to reorganize the data for the SU of the next layer \cite{diana}. In the following explanation, we will refer to the memories preceding and succeeding the reshuffling buffer as the 'before RB' memory and the 'after RB' memory (RB = reshuffling buffer).  

Let's first define the number of data items that are computed in parallel by SU $i$ and that are also required in parallel by SU $j$ of the next layer as $R^{cl}_{i,j}$: 
\begin{dmath}
R^{cl}_{i,j} =  gcd(K_{u,i} \cdot G_{u,i} ,C_{u,j} \cdot G_{u,j}) \cdot gcd(Ox_{u,i},Ox_{u,j}) \cdot gcd(Oy_{u,i},Oy_{u,j})
\end{dmath}
This data cluster of size $R^{cl}_{i,j}$ has to be kept together. 
Based on this, for a set of supported SUs, $R^{cl}_{min}$ is defined as the minimal $R^{cl}_{i,j}$ across all supported ($SU_i$,$SU_j$) combinations. This is the smallest data cluster and hence the smallest reshuffling granularity we have to support. In the example of Fig. \ref{fig:input3}, $R^{cl}_{min} = R^{cl}_{1,2}=2$. 
\stc{Using this concept of data clusters, the number of registers words needed in the reshuffling buffer is
\begin{equation}
REG_{buffer} = \frac{2 \cdot PW_b^{2}}{R^{cl}_{min}}
\label{regbuffer}
\end{equation}
with $PW_b$ the port width of the memories before and after the reshuffling buffer.
The number of MUXes in the reshuffling buffer is equal to
\begin{equation}
MUX_{buffer} \approx PW_b \cdot z\left( \displaystyle\sum \frac{PW_b}{min(PW_b, R^{cl}_{i,j})} \right)
\label{muxbuffer}
\end{equation}
The detailed derivation of both equations can be found in Appendix 3.
}

\subsection{Impact of SU similarity}
One can intuitively feel that configurability overheads will depend on similarities between the different SUs. The challenge is hence to find SUs that are distinct enough to offer utilization benefits across NN layer topologies of interest, yet similar enough to limit the hardware overhead stemming from supporting multiple SUs.
In this subsection, we will therefore assess the impact of the similarity of SUs on the required hardware resources. We will do the evaluation for each of the 3 discussed sources of overhead and show our findings using a numerical example. We will use 4 SUs to illustrate our study: the two we were already using earlier in the example of previous subsections, and two additional SUs, as summarized in Table \ref{table:expSUs}.

\begin{table}[!t]
\renewcommand{\arraystretch}{0.9}
\caption{Used SUs for experiment to show impact of similarity of SUs.}
\vspace*{-0.3cm}
\label{table:expSUs}
\centering
\begin{tabular}{|c|c|}
\hline
SU1 & $[K_u = 2, C_u = 2, Ox_u = 2]$ \\ \hline
SU2 & $[K_u = 2, Ox_u = 4]$ \\ \hline
SU3 & $[G_u = 8]$ \\ \hline
SU4 & $[C_u = 2, Ox_u = 4]$ \\ \hline
\end{tabular}
\vspace*{-0.4cm}
\end{table}

\subsubsection{Data assignment block}
The size of the L1 register is only a function of the highest value of the $A_u$ (Eq. (\ref{au})) and $W_u$ (Eq. (\ref{wu})) across the supported SUs, defined as $A_r$ and $W_r$. The number of MUXes in both stages of the block is given in Table \ref{table:numericalexample}. From this, we see that the number of first stage MUXes does not depend a lot on the similarity of SUs. However, this does change for the number of second stage MUXes. There, similarity in loop unrollings has benefits for the number of MUXes, especially the ones regarding the channels and the $O_{sum}$. 

\begin{table}[!t]
\renewcommand{\arraystretch}{0.9}
\caption{Number of multiplexers in data assignment block for each combination.}
\vspace*{-0.3cm}
\label{table:numericalexample}
\centering
\begin{tabular}{|c|c|c|c|c|c|}
\hline
 & $W_{MUX,1}$ & $A_{MUX,1}$ & $W_{MUX,2}$ & $A_{MUX,2}$ & tot \\ \hline
SU1+SU2 & 4 & 16 & 8 & 8 & 36 \\ \hline
SU1+SU3 & 0 & 8 & 8 & 12 & 28 \\ \hline
SU1+SU4 & 4 & 16 & 8 & 12 & 40 \\ \hline
SU2+SU3 & 4 & 16 & 12 & 8 & 40 \\ \hline
SU2+SU4 & 4 & 24 & 0 & 8 & 36 \\ \hline
SU3+SU4 & 4 & 16 & 12 & 0 & 32 \\ \hline
\end{tabular}
\vspace*{-0.3cm}
\end{table}

\subsubsection{Output aggregation network}
SUs with the same $O_{sum}$ are computed within the same level of the adder tree and therefore don't increase resources, as noted in Eq. (\ref{nadders}). We see that $O_{sum, SU1} = O_{sum, SU4} = 2$ and $O_{sum, SU2} = O_{sum,SU3} = 1$. 
The total number of adders needed is function of the highest $O_{sum}$ from the SUs we want to combine: No adders are needed for the combination of SU2 and SU3, but $N_{adders} = 4$ for every other combination we make with these SUs. 

As illustrated in the results of Table \ref{table:numericalexample_output}, the number of MUXes $O_{MUX}$ in the output aggregation network (Eq. (\ref{eqadder})) depends again on the similarity of the SUs: similarity in $O_{sum}$ drastically reduces overhead in the output aggregation network. 

\begin{table}[!t]
\renewcommand{\arraystretch}{0.9}
\caption{Number of multiplexers in output aggregation network for each combination.}
\vspace*{-0.3cm}
\label{table:numericalexample_output}
\centering
\begin{tabular}{|c|c|c|}
\hline
 & $O_{sum}s$ & $O_{MUX}$  \\ \hline
SU1+SU2 & 2-1 & 12  \\ \hline
SU1+SU3 & 2-1 & 12 \\ \hline
SU1+SU4 & 2-2 & 0 \\ \hline
SU2+SU3 & 1-1 & 8  \\ \hline
SU2+SU4 & 1-2 & 12  \\ \hline
SU3+SU4 & 1-2 & 12  \\ \hline
\end{tabular}
\vspace*{-0.5cm}
\end{table}

\subsubsection{Reshuffling buffer}
Based on the equations for $REG_{buffer}$ (Eq. (\ref{regbuffer})) and $MUX_{buffer}$ (Eq. (\ref{muxbuffer})), the more similar SUs are, the larger $R^{cl}_{min}$ will be, resulting in a smaller reshuffling buffer and less MUXes required, as illustrated in Table \ref{table:numericalexample_reshuffling}.

\begin{table}[!t]
\renewcommand{\arraystretch}{0.9}
\caption{Number of multiplexers in reshuffling buffer for each combination.}
\vspace*{-0.3cm}
\label{table:numericalexample_reshuffling}
\centering
\begin{tabular}{|c|c|c|c|}
\hline
 & $R_{cl}^{min}$ & $REG_{buffer}$ & $MUX_{buffer}$ \\ \hline
SU1+SU2 & 2 & 16 & 12 \\ \hline
SU1+SU3 & 2 & 16 & 12 \\ \hline
SU1+SU4 & 2 & 16 & 12 \\ \hline
SU2+SU3 & 1 & 32 & 28 \\ \hline
SU2+SU4 & 4 & 0 & 0 \\ \hline
SU3+SU4 & 1 & 32 & 28 \\ \hline
\end{tabular}
\vspace*{-0.3cm}
\end{table}

\subsubsection{General insight}
In conclusion, we can say that similarity in many loop parameters of SUs really decreases the resources overhead from configurability. Every type of similarity will impact another part of the flexibility overhead extensions. 
However, to optimize latency and energy it might be interesting to use quite distinct SUs for different layers. This is exactly the trade-off pursued throughout this paper. Section \ref{sectiecoac} will, therefore, introduce an exploration framework, COAC, based on the analytical equations derived above. COAC will be used to perform experiments with real-world architectures and networks in the case study in Section \ref{sectiecasestudy}. 

\begin{figure}[tb]
\centering
\includegraphics[width=0.65\linewidth]{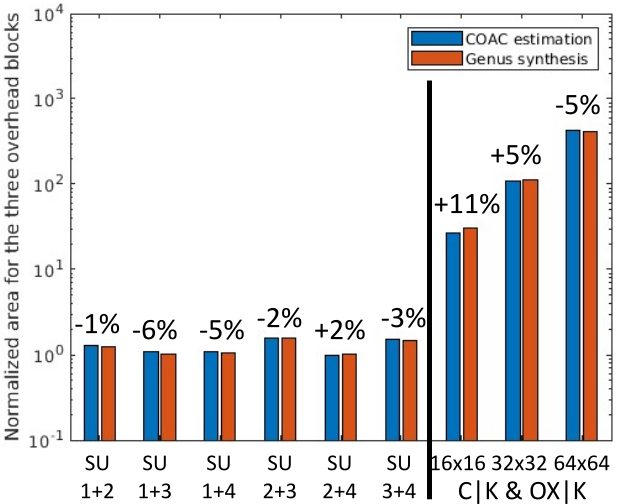}
\caption{\stc{Validation of the overhead area from SU flexibility estimated through COAC, as well as experimentally verified from hardware synthesis. The plotted areas represent the area overhead stemming from the three additional flexibility blocks (Fig. 5).}} 
\label{fig:validation}
\vspace*{-0.5cm}
\end{figure}

\stc{\subsection{Validation}}
\stc{To show the validity of the equations derived in Section III.A-D, we implement the PE interfacing blocks of various combinations of 2 SUs with SystemVerilog and synthesize them with TSMC 16nm standard cell technology. 9 cases were implemented: Firstly, we analyze the 6 cases of the study of Table III that was used to illustrate the impact of the similarity of SUs. These are all quite small architectures but with different SU combining overheads. 
To validate for larger real-world processor implementations, we moreover analyze 3 SU combinations with different PE array sizes. The combined SUs are $C|K$ and $OX|K$, (with $C_u=K_u$, resp. $OX_u=K_u$) as these are very well-known unrollings. The selected PE array sizes are 16x16, 32x32, resp. 64x64. All results are summarized in Fig. \ref{fig:validation}, which depicts for each target implementation the SU flexibility area overhead predicted by the COAC model as well as the actual area overhead derived from the synthesis of the corresponding SystemVerilog implementation. To be transparent to silicon technology impact, we normalize the area to the smallest implementation. It can be observed that the model very closely matches actual silicon implementation results with an error of less than 6\% for 8 out of the 9 cases. 
 Moreover, combinations that have the smallest overhead according to the model, also have the smallest overhead in the actual implementation, proving the usefulness of deploying the model as a design guideline.}

\begin{figure}[tb]
\centering
\includegraphics[width=0.9\linewidth]{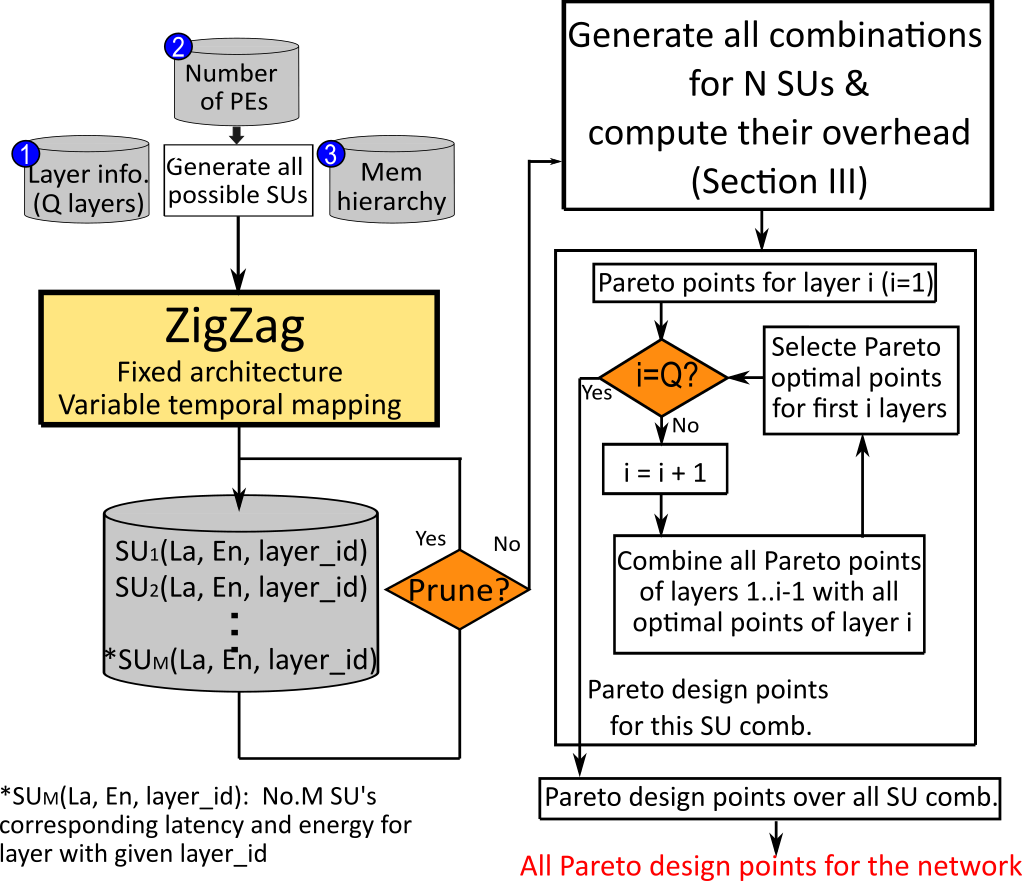}
\caption{COAC flow illustrating mapping space exploration.} 
\label{fig:toolflow}
\end{figure}

\begin{figure}[tb]
\centering
\includegraphics[width=1\linewidth]{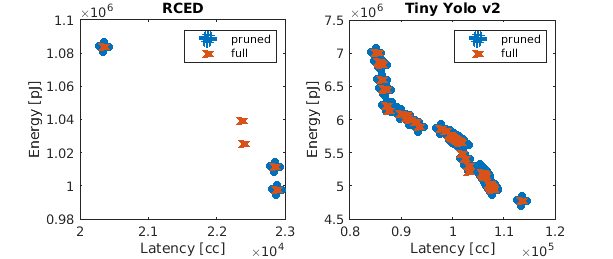}
\caption{Comparison of obtained Pareto curve when using pruned SU set vs every SU. 
} 
\label{fig:pruning}
\vspace*{-0.3cm}
\end{figure}
\vspace*{0.4cm}

\section{COAC: Flexibility-aware design and mapping space exploration tool} 
\label{sectiecoac}
This section illustrates how to find the best combinations of $N$ SUs in an automated way for a given workload or suite of workloads, in terms of area and end-to-end latency and energy. To this end, the developed framework, COAC, explores SU combinations leading to Pareto optimal results in the latency/energy/area domain. 
The framework can be exploited with various values of $N = 1, 2, 3,$ etc. to find the overall optimal solutions. 

Fig. \ref{fig:toolflow} illustrates the toolflow: First, all possible SUs supported by the targeted number of PEs in the array are derived for further investigation. Next, a hardware cost estimation framework, in our studies done through ZigZag \cite{zigzag}, derives for all layers under interest and all possible SUs the temporal unrolling for which latency and energy are minimized (cfr. Section II-C), and their resulting layer execution energy and latency. Subsequently, all possible combinations of $N$ SUs are generated. For each combination, the resources overhead (area) is computed using the equations in Section III. Next, for each of these combinations of $N$ SUs, network level Pareto-optimal design points for latency and energy are derived by iteratively progressing through the different network layers: all energy/latency Pareto points achievable with these $N$ supportive SUs for the first $(i - 1)$ layers are computed, and subsequently combined with all Pareto optimal energy/latency points of the subsequent layer $i$ using one of the SU out of this set of $N$ considered SUs, to derive a set new Pareto-optimal results up to the $i$-th layer. This results in a final set of Pareto-optimal points for the complete end-to-end network execution achievable with this SU combination, and their corresponding values for latency, energy and area. 

In the last step of the COAC flow, all Pareto curves of all SU combinations up to a given maximum number $N$ of SUs to support are put together and the final Pareto optimal solutions are found.
\subsection{SU pruning}
While this flow can be repeated for all SU combinations in a brute force fashion, the user can also activate a pruning step across the SU combinations. In this mode, SUs that have nor the best latency solution for any layer, nor the best energy solution for any layer, are removed before generating SU combinations. This drastically reduces search time, required for processing large networks, while minimally impacting optimality. We illustrate the impact of this SU pruning technique on optimality with the experiment setup from Section V. We assume a 16x16 PE array and therefore 606 possible SUs, according to Eq. (\ref{eq1}). We run COAC both with and without SU pruning for 2 networks (RCED \cite{park2016fully} and Tiny YOLO \cite{yolov2}). Fig. \ref{fig:pruning} shows the Pareto plots for latency/energy for both cases with $N=2$. The difference in found Pareto points is negligible, whereas the pruned approach enables a search time reduction with a factor of more than 1000, which would even increase when combining more SUs. Therefore, in the experiments reported in the case study section, we will adopt COAC with pruning activated. 

\section{Case studies} \label{sectiecasestudy}
\subsection{Experimental setup}
To validate COAC, the benefits and resource overheads of combining two or more SUs for 2 audio processing networks (RCED \cite{park2016fully}, Tiny Yolo\_v2 \cite{yolov2}) and 4 image classification networks (ResNet18, MobileNetv2, Xception \cite{Xception} and VGG19 \cite{VGG19} for the ImageNet dataset \cite{imagenet}) are estimated and analyzed. An accelerator architecture template with the same memory hierarchy as Evolver \cite{evolver} (a 256KB weight buffer with 4096b port width and a 156KB activation buffer with 1024b port width) is assumed in TSMC 16nm. 

The architecture contains 16x16 = 256 PEs. In the first step, COAC determines all valid SUs that can be evaluated on this PE array, that satisfy Eq. \ref{eq1} for $nb_{PEs} = 256$. 

We will first assess the achievable energy/latency/cost trade-offs when selecting optimal sets of $N=$1, 2 and 3 SUs. Next, the obtained solutions are compared with the SUs supported in the original Evolver architecture \cite{evolver}, which supports 6 different SUs: $[C_u = 1/2/4/8/16/32]$, and corresponding $[K_u =256/128/64/32/16/8]$, resp. \stc{The Evolver overhead stemming from the support of the different SUs can be modelled using the same three hardware overhead blocks modeled in COAC.}  \stc{also the off-chip memory accesses are modeled, by describing the off-chip memory as a ‘top level memory’ in ZigZag, with following bandwidth and energy per read/write access:}

Research questions we want to answer are:
\begin{itemize}
    \item What are latency/energy benefits when combining SUs? 
    \item How many SUs are optimally combined? 
    \item What is the corresponding resource overhead? Is this overhead a function of the combined SUs?
    \item Are the SUs used by Evolver close to optimal SUs in terms of latency/energy/resources overhead, or far away from them? This would allow to show the importance of a tool like COAC to find optimal mapping combinations. 
    \item What's the impact of selecting more SUs when considering more workloads in parallel?
\end{itemize}
\stc{The total COAC simulation time for all performed experiments was 3 days.}

\subsection{Results}
\subsubsection{Latency and energy gains}
\begin{figure}[tb]
\centering
\includegraphics[width=1.0\linewidth]{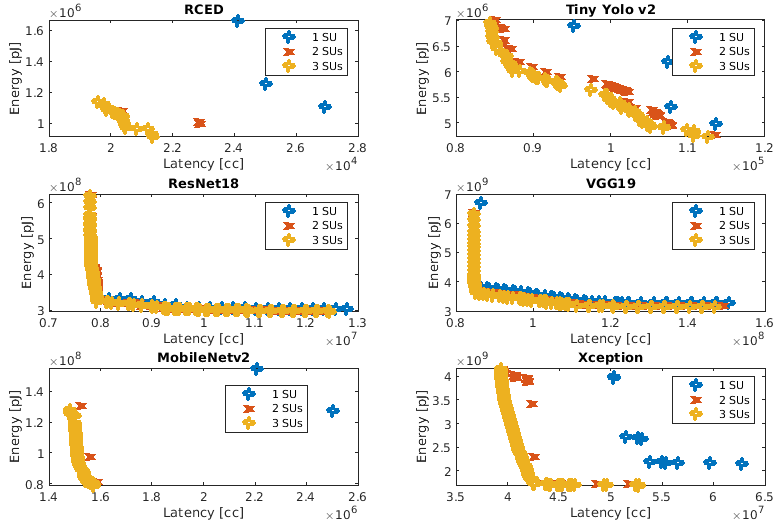}
\caption{Pareto plots for latency/energy when combining 1/2/3 SUs. The networks are optimized separately.} 
\label{fig:allresults}
\vspace*{-0.3cm}
\end{figure}
Fig. \ref{fig:allresults} shows the Pareto curves in the latency/energy space for $N = $ 1, 2 and 3 SUs for the six different networks. As we can see, for all networks except VGG19 and ResNet18, combining 2 SUs clearly outperforms using only a single SU for both latency and energy. Layers in VGG19 and ResNet18 are all quite similar (same kernel sizes in terms of Fx, Fy and G), resulting in a single optimal SU. For all other networks, the variability in layer types is abundant enough to justify a need for 2  supported SUs. A deeper analysis reveals that for a network with both depthwise and classical layers, such as Xception or MobileNetv2, COAC tunes 1 of the combined SUs to have $G_u > 1$ and the other SU dedicated to $G_u = 1$. For RCED and Tiny Yolo, COAC exploits the difference in kernel size, in the selected SUs with different supported $Fx_u$, as this is here the most varying loop dimension. 
We notice that the benefits for going from 1 to 2 SUs are highest for MobileNetv2. Here, the optimal EDP (energy-delay-product) is reduced with 59.5\%. 
When going from 2 to 3 SUs for a single network, the Pareto curve does not shift a lot anymore, while increasing resource overhead. 

Let's now analyze what happens if we want to optimize the supported set of SUs for all networks together. This is done by entering all 6 networks as the target workload in COAC. As illustrated in Fig.  \ref{fig:allresults}, not all networks take the same order of magnitude in execution cost. Therefore, we perform a preprocessing step to make all networks roughly equally important, by normalizing all network performances to their most optimal execution cost: Specifically, for each network the best individual SU is computed in terms of latency. The latency for this SU for the complete network is called $L_{best}$. Afterwards, for each layer and each investigated SU, both latency and energy are divided by $L_{best}$. This equalizes the impact of all networks to the experiment, and makes that the optimal 'relative latency' for a complete network is equal to 1. 
\begin{figure}[tb]
\centering
\includegraphics[width=0.8\linewidth]{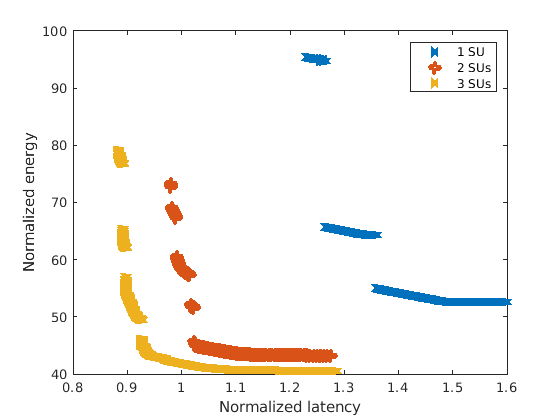}
\caption{Pareto plots for latency/energy when combining 1/2/3 SUs of a common accelerator platform required to support and execute all networks. The values for normalized latency and energy are divided by 6 to have mean values for 1 network.} 
\label{fig:allresultsshared}
\vspace*{-0.3cm}
\end{figure}
The results are summarized in Fig. \ref{fig:allresultsshared}. We see that the Pareto fronts continue shifting to the left bottom when adding a third SU. We have a 38\% EDP gain when going from 1 to 2 SUs and a further 12\% gain if we combine 3 SUs. This was not the case when we assessed the optimal SU combinations for each network individually, were no benefits were observed beyond sets of 2 SUs. This makes sense, as the more diversity in the layers to be executed, the harder it becomes to efficiently execute them with a limited set of SUs. The EDP gains of this combined SU optimization, and summarized in Table \ref{table:gainspernetwork}, split out for each network independently. The best EDPs for both 1 SU and combination of 2 SUs are taken. 
We see that the highest gain is again for MobileNetv2.  However, it is slightly reduced to 50\% gain, where it was 59.5\% if we could optimize for MobileNetv2 only. This matches with intuition, as we are jointly optimizing for all networks instead of for a single one. 

\begin{table}[!t]
\renewcommand{\arraystretch}{1.2}
\caption{EDP gains from 1 to 2 SUs for each network separately using same set of shared-optimized SUs. 
}
\vspace*{-0.3cm}
\label{table:gainspernetwork}
\centering
\begin{tabular}{|c|c|c|c|}
\hline
Network & \begin{tabular}{c} $1 SU$ \\  COAC \end{tabular} &\begin{tabular}{c} $2 SUs$ \\  COAC \end{tabular} & Evolver \cite{evolver} \\ \hline
RCED & base & -35\% &  -77\% \\ \hline
Tiny Yolo v2 & base & -1\% & -47\% \\ \hline
ResNet18 & base & -34\% &  -33\% \\ \hline
VGG19 & base & -12\% &  -3\% \\ \hline
MobileNetv2 & base & -50\% & -14\% \\ \hline
Xception & base & -38\% & -60\% \\ \hline
ALL TOGETHER & base & -38\% & -45\%\\ \hline
\end{tabular}
\vspace*{-0.3cm}
\end{table}

\subsubsection{Selected spatial unrollings}  
Let's now look into more detail to which SU combinations are retained by the COAC optimization in the previous multi-network study 
of Fig. \ref{fig:allresultsshared}. In this discussion, we take the SU(s) on the energy-latency Pareto plot for which the EDP value is lowest. If only one SU can be used, COAC picks $[Ox_u = 16, K_u = 16]$. If two SUs can be combined, COAC picks $[Ox_u = Oy_u = 4, K_u = 16]$ and $[Ox_u = 16, Fx_u = 4, K_u = 4]$. If 3 SUs can be combined, the tool picks $[Ox_u = Oy_u = 4, K_u = 16]$, $[Ox_u = 16, C_u = 2, K_u = 8]$ and $[Ox_u = 32, Fx_u = 4, G_u = 2]$. 

These sets explain why we have benefits from combining SUs. All picked SUs unroll spatially over $Ox$ as that's the only loop dimension available in all layers across all networks (classical 3x3, pointwise, depthwise). If only one SU is used, this is combined with unrolling over K to minimize output bandwidth requirements. When we combine 2 SUs, we can pick one with unrolling over $Fx$ for data reuse and resulting memory access reasons, and one without unrolling over $Fx$ for the pointwise layers. A spatial unrolling over $G$ is not included yet, as apparently for this set of networks with according relative importance, the impact of the depthwise layers is not large enough. That's why we still have benefits from combining with a third SU: we can now add one where unrolling over $G$ is included. \stc{Yet, in our experiments, this addition of the third SU only made sense for networks that exploit depthwise layers like MobileNetv2 and not for networks like ResNet18.} The resources that are necessary to implement these combinations of SUs will be discussed in the next section, where we compare the required area with the area in Evolver.

\section{State-of-the-art comparison}
\subsection{Quantitative comparison with Evolver}
\begin{figure}[tb]
\centering
\includegraphics[width=0.9\linewidth]{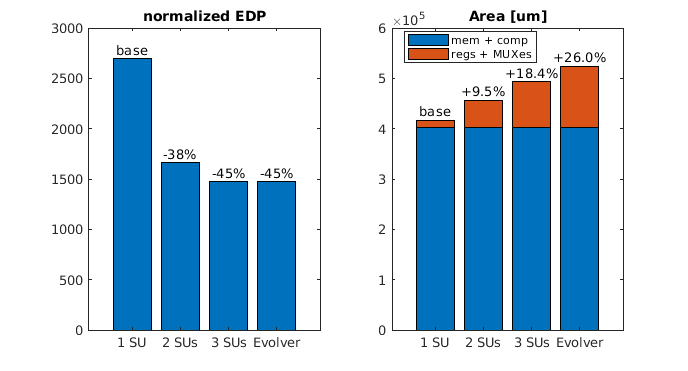}
\caption{Comparison of Evolver architecture with our approach.  
} 
\label{fig:evolver}

\end{figure}

Finally, we can compare the best SU combinations found, with the SU combinations supported in a flexible architecture such as Evolver \cite{evolver}. Fig. \ref{fig:evolver} compares results achievable with Evolver's 6 supported SUs with the results obtained by COAC. We see that the EDP of Evolver is equally good as our approach with only 3 SUs, where Evolver needs 6. Therefore, the COAC optimized architecture with only 3 SUs has much fewer resources than the Evolver architecture, as can be seen in the second subplot, in which all designs are normalized to the same 16nm technology. 
The area is here depicted in two parts: the blue part contains the memory and computation blocks, the orange part contains the area to support unrolling flexibility in terms of registers and multiplexers. Evolver clearly needs more area than our found solution, which proves COAC is an interesting tool to find efficient solutions for area constrained solutions. Interesting to note is also that Evolver has fewer resources than if the trend of the '1 SU', '2 SUs' and '3 SUs' would be followed. This is due to the fact that the SUs in Evolver only vary in unrolling in C and K. This is a practical proof that similarity in SUs indeed makes sense for the resource overhead, as already derived mathematically in this manuscript. \stc{When multiplying area with EDP, the solution found by COAC clearly outperforms the Evolver solution, a similar EDP axchieved for lower area.} A numerical analysis per network separately is also given in Table \ref{table:gainspernetwork}. 

\subsection{Qualitative comparison with other frameworks}
COAC is the first framework to support the assessment of the best unrolling combinations across complete neural networks, taking energy, latency and cost jointly into account. These capabilities are currently lacking in state-of-the-art hardware design exploration frameworks in literature, such as ZigZag \cite{zigzag}, Timeloop \cite{timeloop} and Union \cite{union}, as summarized in Table \ref{table:sota}. In the introduction of this paper, we talked about 3 different approaches to map a convolutional network on a PE array: using a fixed SU for all layers (lowest flexibility), using an optimized SU for each individual layer (highest flexibility) and a hybrid version with a limited number of SUs for the complete network. Existing SotA design exploration frameworks are only able to evaluate the first two cases, whereas COAC is able to do all of them. Moreover, COAC also analyzes the hardware overhead that comes with this flexibility. Fig. \ref{fig:evolver} has shown quantitatively that COAC is able to boost performance in terms of latency and energy in comparison to SUs published in literature. 
 
\begin{table}[!t]
\renewcommand{\arraystretch}{1.2}
\caption{Comparing COAC to ZigZag \cite{zigzag}, Timeloop \cite{timeloop} and Union \cite{union}. }
\vspace*{-0.15cm}
\label{table:sota}
\centering
\begin{tabular}{|c|c|c|c|c|}
\hline
 &  ZigZag  & Timeloop  & Union & \textbf{COAC} \\ \hline
1 SU per network  &  yes & yes & yes &  \textbf{yes} \\ \hline
1 SU per layer  &  yes & yes & yes &  \textbf{yes} \\ \hline
Few SUs per network  &  no & no & no & \textbf{yes} \\ \hline
\begin{tabular}{@{}c@{}} Overhead of combining SUs \end{tabular}  &  no & no & no & \textbf{yes} \\ 
\hline
\end{tabular}
\vspace*{-0.3cm}
\end{table}

\section{Conclusion}
Different neural network layer topologies benefit from different hardware parallelization dimensions, denoted by the Spatial Unrolling (SU) parameters. It is therefore beneficial to support multiple SUs, ideally in the same PE array in order to save area. This work presents COAC, a framework to find the optimal set of SUs to minimize latency/energy for end-to-end inference, as well as resource overhead that comes from the introduced flexibility. Three required architecture extensions lead to this resource overhead. The first is a flexible data assignment block, to make sure the correct weights and data activations go to the correct PE element for each SU. The second extension has to do with the flexible data aggregation network. In this flexible adder tree, the outputs belonging to each SU might be computed at different levels of the adder tree, leading to the need for multiplexers to select the correct output to store in the memory. The third extension is a reshuffling buffer. Not all inputs that are needed in parallel for a given SU are computed in parallel by the previous SU. Therefore, the reshuffler makes sure that these data items are stored together in the memory to reduce the number of stalling cycles as much as possible. 

We model the cost of each of these blocks in function of the supported SU set and showed analytically the impact of similarity of SUs. The more similar the SUs are, the smaller the resources overhead will be. Subsequently, a search procedure to find the most energy/latency optimal set is integrated in a new automated exploration framework, named COAC, together with a pruning method to reduce search time. COAC shows to find better SU combinations than those used in literature. 
The benefits for combining SUs differs from network to network: the more diversity in layer types (e.g. both classical and depthwise layers), the more beneficial it is to combine SUs in terms of latency/energy. 

The code of this project will be made publicly available on www.github.com/StevenColleman1234. 




\begin{thebibliography}{00}
\bibitem{resnet}
He, Kaiming, et al. "Deep residual learning for image recognition." Proceedings of the IEEE conference on computer vision and pattern recognition. 2016.

\bibitem{mobilenetv2}
Sandler, Mark, et al. "Mobilenetv2: Inverted residuals and linear bottlenecks." Proceedings of the IEEE conference on computer vision and pattern recognition. 2018.

\bibitem{mobilenet}
Howard, et al. "Mobilenets: Efficient convolutional neural networks for mobile vision applications." arXiv preprint arXiv:1704.04861.

\bibitem{attention}
Vaswani, Ashish, et al. "Attention is all you need." Advances in neural information processing systems 30 (2017).

\bibitem{heterogeneous}
Kwon, H., et al. "Heterogeneous Dataflow Accelerators for Multi-DNN Workloads." 2021 IEEE International Symposium on High-Performance Computer Architecture (HPCA). IEEE, 2021.

\bibitem{loma}
Symons, A., et al, 2021, June. LOMA: Fast Auto-Scheduling on DNN Accelerators through Loop-Order-based Memory Allocation. In 2021 IEEE 3rd International Conference on Artificial Intelligence Circuits and Systems (AICAS) (pp. 1-4). IEEE.

\bibitem{eyeriss}
Chen, Y., et al. "Eyeriss v2: A flexible accelerator for emerging deep neural networks on mobile devices." IEEE Journal on Emerging and Selected Topics in Circuits and Systems 9.2 (2019): 292-308.

\bibitem{zigzag}
Mei, L., et al. "ZigZag: A memory-centric rapid DNN accelerator design space exploration framework." arXiv preprint arXiv:2007.11360 (2020).

\bibitem{timeloop}
Parashar, A., et al. "Timeloop: A systematic approach to dnn accelerator evaluation." 2019 IEEE international symposium on performance analysis of systems and software (ISPASS). IEEE, 2019.

\bibitem{Maestro}
Kwon, H., et al. 2020. Maestro: A data-centric approach to understand reuse, performance, and hardware cost of dnn mappings. IEEE micro, 40(3), pp.20-29.

\bibitem{Interstellar}
Yang, Xuan, et al. "Interstellar: Using halide's scheduling language to analyze dnn accelerators." Proceedings of the Twenty-Fifth International Conference on Architectural Support for Programming Languages and Operating Systems. 2020.


\bibitem{tpu}
Jouppi, Norman P., et al. "In-datacenter performance analysis of a tensor processing unit." Proceedings of the 44th annual international symposium on computer architecture. 2017.



\bibitem{park2016fully}
Park, Se Rim, et al. "A fully convolutional neural network for speech enhancement." arXiv preprint arXiv:1609.07132 (2016).

\bibitem{yolov2}
Fanioudakis, et al. "Deep networks tag the location of bird vocalisations on audio spectrograms." arXiv preprint arXiv:1711.04347 (2017).

\bibitem{Xception}
Chollet, F., 2017. Xception: Deep learning with depthwise separable convolutions. In Proceedings of the IEEE conference on computer vision and pattern recognition (pp. 1251-1258).

\bibitem{VGG19}
Simonyan, K. , et al. 2014. Very deep convolutional networks for large-scale image recognition. arXiv preprint arXiv:1409.1556.


\bibitem{imagenet}
J. Deng et al., “Imagenet: A large-scale hierarchical image database,” in 2009 IEEE conference on computer vision and pattern recognition. Ieee, 2009, pp. 248–255.



\bibitem{evolver}
Tu, F., et al. "Evolver: A deep learning processor with on-device quantization–voltage–frequency tuning." IEEE Journal of Solid-State Circuits 56.2 (2020): 658-673.

\bibitem{union}
Jeong, Geonhwa, et al. "Union: A Unified HW-SW Co-Design Ecosystem in MLIR for Evaluating Tensor Operations on Spatial Accelerators." 2021 30th International Conference on Parallel Architectures and Compilation Techniques (PACT). IEEE, 2021.

\bibitem{spatially}
Colleman, Steven, et al. "Processor Architecture Optimization for Spatially Dynamic Neural Networks." 2021 IFIP/IEEE 29th International Conference on Very Large Scale Integration (VLSI-SoC). IEEE, 2021.

\bibitem{highutilisation}
Colleman, Steven, et al. "High-utilization, high-flexibility depth-first cnn coprocessor for image pixel processing on fpga." IEEE Transactions on Very Large Scale Integration (VLSI) Systems 29.3 (2021): 461-471.

\bibitem{Flagship}
J. -W. Jang et al., "Sparsity-Aware and Re-configurable NPU Architecture for Samsung Flagship Mobile SoC," 2021 ACM/IEEE 48th Annual International Symposium on Computer Architecture (ISCA), 2021, pp. 15-28, doi: 10.1109/ISCA52012.2021.00011.

\bibitem{diana}
K. Ueyoshi et al., "DIANA: An End-to-End Energy-Efficient Digital and ANAlog Hybrid Neural Network SoC," 2022 IEEE International Solid- State Circuits Conference (ISSCC), 2022, pp. 1-3, doi: 10.1109/ISSCC42614.2022.9731716.

\bibitem{us}
Colleman, Steven, et al. "Optimizing Accelerator Configurability for Mobile Transformer Networks." 2022 IEEE 4th International Conference on Artificial Intelligence Circuits and Systems (AICAS). IEEE, 2022.

\bibitem{edge}
Yazdanbakhsh, Amir, et al. "An evaluation of edge tpu accelerators for convolutional neural networks." arXiv preprint arXiv:2102.10423 (2021).

\end{thebibliography}
\section*{Acknowledgment}
This project has been partly funded by the European Research Council (ERC) under grant agreement No. 101088865, the European Union’s Horizon 2020 programme under grant agreement No. 101070374, the Flanders AI Research Program and KU Leuven.

\section*{\stc{Appendix 1: Data assignment block cost analysis}}
First, we will analyze the first stage of MUXes, which selects the relevant section of the L2 weight and activation memory words. We start our analysis with the weights part. This set of MUXes will ensure that the $W_u$ weights for a given SU are always buffered at the $W_u$ leftmost (lowermost in the figure) positions of the L1 register. In the example of Fig. \ref{fig:input3} these are the $W_{u,SU1} = 4$ lowermost (= all) positions for SU1 and $W_{u,SU2} = 2$ lowermost positions for SU2. 
To still ensure full utilization of the L2 memories with $PW_{L2,weights} > W_u$ (PW = port width = physical width of the memory in number of words) for at least one SU, MUXes are required in the first stage, as different data words from the same L2 weight memory address will have to go to the same position in the L1 register across different supported SUs. 

For each register position $i$, the number of MUXes in front of the L1 register depends on the smallest $W_u$ from all different SUs that need this register position $i$. 
In the example of Fig. \ref{fig:input3}, weight L1 registers $W_1$ and $W_2$ have to store weights for both SU1 and SU2, hence 
weights for $W_1$ and $W_2$ can come from 2 different positions and therefore need 2 '1-input MUXes' in front of them. Positions $W_3$ and $W_4$ only have to store weights for SU1. As $PW_{L2,weights} = W_{u,SU1}$, no MUXes are needed here. In general terms, the total number of one-input MUXes in the first stage for the weights assignment is
\begin{equation}
W_{MUX,1} = \displaystyle\sum_{i=1}^{W_r} z \left( \ceil*{ \frac{PW_{L2,weights}}{min(W_u), \forall \ SU_j | (W_{u,SU_j} \ge i)}} \right)
\end{equation}
In this, $z(x)$ is a function that returns 0 if $x = 1$ and $x$ itself otherwise.  

Let's now switch the attention to the first MUX stage for the activation data, feeding the L1 activation registers. Here, for the different SUs data has to be arranged taking into account padding and convolutional input data reuse, hence requiring that every register word can access every L2 activation memory position that contains data of the same input channel. Therefore, for each activation register $A_i$, the data can come from $\frac{PW_{L2,act}}{G_u.C_u}$ positions for a given SU if the value of $A_u$ for this SU is at least $i$. This stems from the assumption that the $PW_{L2,act}$ activations at any L2 address belong to $G_u.C_u$ different input channels and we only need to reshuffle the data for a given channel, not across channels. 

For supporting multiple SUs, we have to take the minimal value for $G_u.C_u$ across all supported SUs, as this will lead to more possible connections and therefore more MUXes.  
As such, we find that this requires following number of one-input MUXes in the first activation stage:
\begin{equation}
A_{MUX,1} = \displaystyle\sum_{i=1}^{A_r} z\left( \ceil*{\frac{PW_{L2,act}}{ min(G_u \cdot C_u), \forall \ SU_j | (A_{u,SU_j} \ge i)}}\right)
\end{equation}
hence instantiating MUXes for each register $i$, supporting the input locations of all unrollings for which $A_{u,SU_j} \ge i$. If only SU1 would be supported in the example of Fig. \ref{fig:input3}, we would have enough with 2 inputs in front of each register. That is because $C_{u,SU1} = 2$ and therefore only $\frac{PW_{L2,act}}{G_u.C_u}=2$ activations from a given L2 activation address belong to the same channel. But we also have to support SU2, and as $C_{u,SU2} = 1$, all activations in L1 can come from all places in L2. 

\begin{figure}[tb]
\centering
\includegraphics[width=1\linewidth]{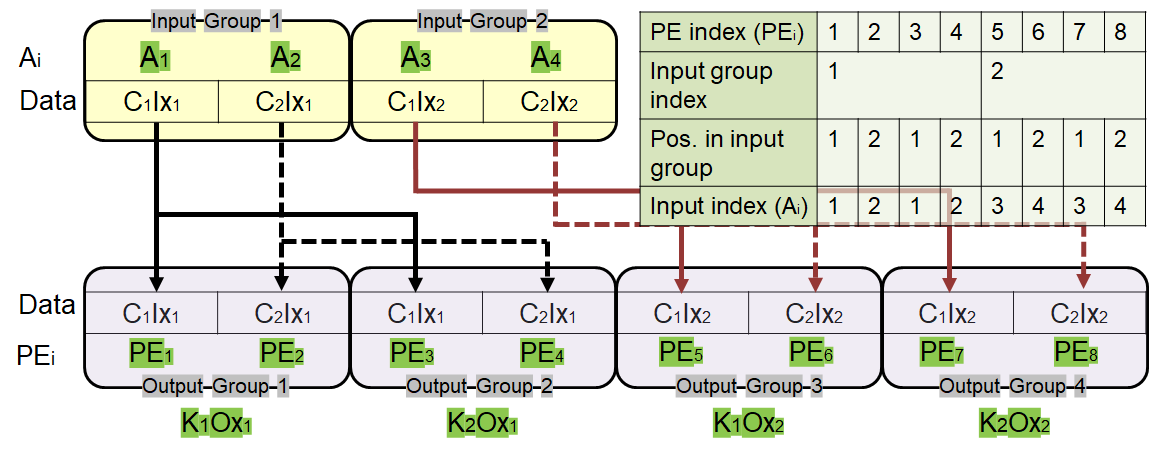}
\caption{Illustration for the number of MUXes in the second stage, with SU1 as example.} 
\label{fig:Aij}
\vspace*{-0.5cm}
\end{figure}

Finally, we continue with the analysis of the second set of MUXes, which sits between the L1 registers and the PE array. We start with the activations MUXes. For a given SU $j$, our task is to derive the register words $A(i)$, that have to be sent as input to $PE_i$, as illustrated for SU1 in Fig. \ref{fig:Aij}. In this reasoning, we will exploit the memory data organization for inputs and outputs, as mentioned in Section III-A. 
To assess this analytically, we divide both the L1 and the PEs in groups of size $O_{sum}$ (as $O_{sum}$ subsequent activations from L1 have to go to $O_{sum}$ subsequent PEs), which is 2 in the case of Fig. \ref{fig:Aij} and call these "input groups" of the MUX stage (for L1) and "outputs groups" of the MUX stage (for the PE array), respectively. As shown in Fig. \ref{fig:Aij}, the first $K_u$ output groups (output group 1 and output group 2) require data from input group 1 (A1 and A2 registers), one input group is shared among these two output groups, as they perform processing with the same inputs for a different output channel. The next $K_u$ output groups (output group 3 and output group 4) require activations from input group 2 as they perform processing for a different (Ox,Oy) position, hence with different inputs. In general, every $K_u$ output groups (a.k.a $K_u*O{sum}$ PEs) need the same input group of data. Therefore, $PE_i$ requires data from input group:
\begin{equation}
input_{group,PE_i} = \ceil*{\frac{i}{K_u \cdot O_{sum}}}
\label{eqinput}
\end{equation}

As an example, the corresponding input group index of each PE of Fig. \ref{fig:input3} is illustrated in the second row of the table in Fig.~\ref{fig:Aij}. Now, we have to derive the relative position of $PE_i$ within the input group, which is a value in the range $[1, O_{sum}]$. $PE_1$ until $PE_{O_{sum}}$ need positions $1$ until $O_{sum}$ in the first input group, $PE_{O_{sum} + 1}$ until $PE_{2 \cdot O_{sum}}$ need positions $1$ until $O_{sum}$ in the first input group as well, $PE_{2 \cdot O_{sum} + 1}$ until $PE_{3 \cdot O_{sum}}$ need positions $1$ until $O_{sum}$ in the second input group and so on, or: 
\begin{equation}
position_{group,PE_i} = mod(i-1,O_{sum}) + 1
\end{equation}
Or equivalently: 
\begin{equation}
position_{group,PE_i} = i - \left( \ceil*{\frac{i}{O_{sum}}} - 1 \right ) \cdot O_{sum} 
\label{eqposition}
\end{equation}
This number is illustrated at the third row of the table in the example of Fig. \ref{fig:Aij}. Based on Eq. (\ref{eqinput}) and Eq. (\ref{eqposition}), $A(i)$, the index of the source input activation register for $PE_i$ for for a specific SU can be computed as: 
\begin{equation}
A(i) = \left( input_{group,PE_i} - 1 \right) \cdot O_{sum} + position_{group,PE_i}
\label{eqA1}
\end{equation}
which leads after simplification to
\begin{equation} 
A(i) = i - \left( \ceil*{\frac{i}{O_{sum}}} - \ceil*{\frac{i}{K_{u}  \cdot  O_{sum}}} \right) \cdot O_{sum}
\label{temp}
\end{equation}
These indices are denoted in the last row in the example of Fig.~\ref{fig:Aij}.
We can easily see that this equation is also valid for SUs with an unrolling in $G_u$. Using the equation for $A_u$, we see that $A_u$ is equal to the number of PEs if $K_u = 1$. This means that $PE_i$ needs an activation from register $i$ in the L1 activation register. 
Indeed, if we fill in $K_u = 1$ in Eq. \ref{temp}, the part between brackets becomes 0 such that we remain $A(i) = i$ and the equation holds.

Therefore, in general terms, when executing SU $j$, $PE_i$ needs to source input activation data from register word $A(i,j)$ where
\begin{equation} 
A(i,j) = i - \left( \ceil*{\frac{i}{O_{sum,j}}} - \ceil*{\frac{i}{K_{u,j}  \cdot  O_{sum,j}}} \right) \cdot O_{sum,j}
\label{formulaAij}
\end{equation}
The number of different values in $A(i,:)$ across SUs indicates the number of different positions from which activation data for $PE_i$ might come across the different supported SUs, hence allowing to analytically compute the number of MUXes required for a given combination of SUs.\\ 
We now want to perform a similar reasoning for the weights, were we will use the notation $W(i,j)$. Due to the order of the computations in the PE (to store the output data in a $Oy_u|Ox_u|K_u$ or $Oy_u|Ox_u|G_u$ way, the first $W_u$ PEs compute all channel outputs for a given output position $(Ox, Oy)$, the next $W_u$ PEs perform computations for all channel outputs for a next output position and so on. 
This means that the first $W_u$ PEs need weights from $W_1$ until $W_u$ and every next $W_u$ PEs need the same $W_u$ weights. As such, we can express $W(i,j)$, the register address sourcing $PE_i$ under SU $j$, as:
\begin{equation}
W(i,j) = mod(i-1,W_{u,j}) + 1
\end{equation}
This again allows to compute the number of different weight source registers per PE, and hence the number of required multiplexers between the weight registers and the PE array.

\section*{\stc{Appendix 2: Output aggregation network cost analysis}}
\begin{figure}[tb]
\centering
\includegraphics[width=0.65\linewidth]{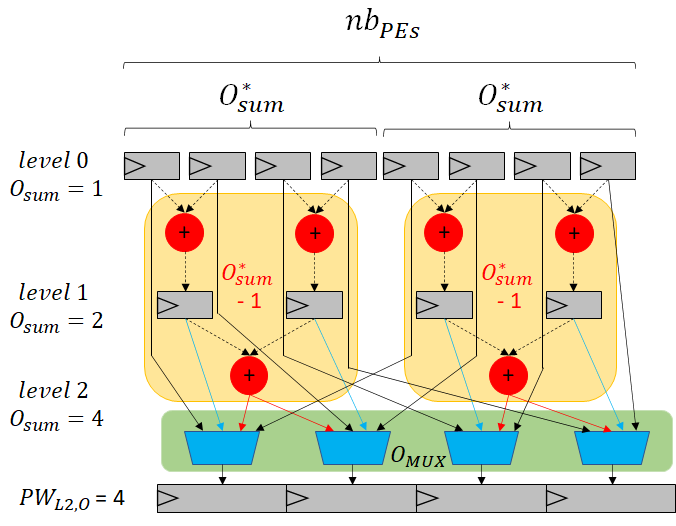}
\caption{Illustration for the number of MUXes in output aggregation network.} 
\label{fig:adderuitleg}
\vspace*{-0.5cm}
\end{figure}

MUXes are required to extract accumulation results from various levels in the adder tree, to store them into the L1 output registers in function of the SU under execution. 
The total number of MUXes needed, can be found by assessing the different levels of the adder trees for their consumption profiles. The output of level $i$ in the tree is defined as the level where $2^i$ intermediate results are added. 
E.g. in our example of Fig. \ref{fig:adderuitleg}, the adder tree has a level 0 (final output for SU where $O_{sum} = 1 = 2^0$), a level 1 (final output for SU where $O_{sum} = 2 = 2^1$), etc.

\subsubsection{Within one SU}
If level $i$ contains final outputs, then there are $\frac{nb_{PEs}}{2^i}$ outputs from different adders that have to be written to the lowest level output features memory L2.
The port width of this memory is defined as $PW_{L2,O}$ (expressed in number of words). 
If $\frac{nb_{PEs}}{2^i} \leq PW_{L2,O}$, output data at a given position for the output memory can only come from one adder of the level. For level 1 in the example of Fig. \ref{fig:adderuitleg}, $8/2^1 = 4 = PW_{L2,O}$ and therefore, each of the 4 positions of the L2 output memory can get data from one adder's result in level 1. Simultaneously, each of the 4 positions of the L2 output memory can also only get data from one adder's output of level 2. If $\frac{nb_{PEs}}{2^i} > PW_{L2,O}$, the data will have to be written back in memory in multiple write cycles and hence each position of the L2 memory can get data from $\frac{\frac{nb_{PEs}}{2^i}}{PW_{L2,O}}$ adder's results. As shown in Fig.~\ref{fig:adderuitleg}, this leads to 2 possible L2 input sources from level 0. 

\subsubsection{Across supported SUs}
Across the different SUs, the total number of adders from which each position of the L2 memory can get data from is the sum of what is required from each level of the adder tree used for final outputs of at least 1 SU. In our example of Fig.~\ref{fig:adderuitleg}, this amounts to 1 (from level 2) + 1 (from level 1) + 2 (from level 0) = 4 and therefore 4 one-input MUXes are needed between the PE array and the L2 memory.

In general, the total number of one-input MUXes ($O_{MUX}$) in the output aggregation network can be formulated as:
\begin{equation}
O_{MUX} = PW_{L2,O} \cdot z\left( \displaystyle\sum_{i} f(i) \cdot max \left(\frac{\frac{nb_{PEs}}{2^i}}{PW_{L2,O}}, 1 \right) \right)
\end{equation}
where $f(i)$ is 1 if level $i$ contains final outputs for at least 1 of the supported SUs and 0 otherwise. The $max \left(\frac{\frac{nb_{PEs}}{2^i}}{PW_{L2,O}}, 1 \right)$ stems from the fact that for the highest number of levels, output data at a given memory position can only come from 1 adder output in the adder tree whereas for the lower number levels, this number of adder outputs is equal to $\frac{\frac{nb_{PEs}}{2^i}}{PW_{L2,O}}$. 
The total number of MUXes per output memory position has to be multiplied with the total number of words in the output memory port $PW_{L2,O}$.

\section*{\stc{Appendix 3: Reshuffling buffer overhead modeling}}
\subsection{Reshuffling buffer size}
Define $PW_b$ as the port widths of the memories before and after the reshuffling buffer, as depicted in Fig. \ref{fig:input3}, expressed in number of data words that can be sent/received during one cycle. Note that our tool can also support the analysis where these port widths are not equal but we will here assume here they are equal to make the reasoning easier to follow. If $R^{cl}_{min}$ is a multiple of $PW_b$, we can split the data to several $PW_b$ sized clusters and directly send them out cycle by cycle to the 'after RB' memory. Therefore, no reshuffling buffer is needed at this condition. 

When $R^{cl}_{min}$ is not a multiple of $PW_b$, an actual physical instantiation of a reshuffling buffer will be necessary. In the example of Fig.\ref{fig:input3}, for the transfer from SU1 to SU2, the reshuffling buffer has to fetch $PW_b/R^{cl}_{1,2} = 2$ data clusters per clock cycle ($PW_b$ is 4 in this example while $R^{cl}_{1,2}$ equals 2) from the 'before RB' memory to perfectly eliminate the input access stalling. In Fig.\ref{fig:input3}, the data belonging to the same cluster is indicated in red and green, respectively. Each of these clusters must be placed at a different memory address in the 'after RB' memory. To enable this, the data belonging to the same output word is loaded over consecutive clock cycles into the flexible reshuffling buffer.
As only $R^{cl}_{1,2}$ data from the same data cluster are read from the 'before RB' memory every clock cycle, it takes $PW_b/R^{cl}_{1,2}$ clock cycles to obtain all data from the 'before RB' memory that we want to write at the same address of the 'after RB' memory. During these cycles, a total number of $\frac{PW_b}{R^{cl}_{1,2}} \cdot PW_b$ words hence are read (8 words in the example of Fig.\ref{fig:input3}). The reshuffling buffer must be large enough to store all the data read from the 'before RB' memory during these cycles. 
In addition, the reshuffling buffer should to be designed as a double buffer to avoid data stalls while a batch of data is read into the 'after RB' memory. Therefore, the number of register words needed in the reshuffling buffer is in general: 
\begin{equation}
REG_{buffer} = \frac{2 \cdot PW_b^{2}}{R^{cl}_{min}}
\end{equation}

\subsection{Number of MUXes for reshuffling buffer}
There is no need for MUXes between the 'before RB' memory and the reshuffling buffer, as the reshuffling buffer just contains multiple concatenated 'before RB' memory lines. However, there are MUXes needed between the reshuffling buffer and the 'after RB' memory to concatenate the correct data clusters for parallel writing to the 'after RB' memory.
For the two supported SUs (SU1 to SU2) in Fig. \ref{fig:input3}, 4 data access patterns should be enabled to cover the various SUs execution orders: SU1 to SU1, SU2 to SU1, SU1 to SU2 and SU2 to SU2. Based on the analysis conducted in above, in this example all data for the 'after RB' memory can come from 2 $(= PW_{b}/R^{cl}_{1,2})$ places in the reshuffling buffer. As $R^{cl}_{1,1} = R^{cl}_{2,1} = R^{cl}_{2,2} = 4 = PW$, the transformation from SU1 to SU1, SU2 to SU1 and SU2 to SU2 will not need intermediate buffering for the reshuffling, since a memory line from the 'before RB' memory encompasses a complete data cluster and can hence be written immediately into the 'after RB' memory. Only SU1 to SU2 requires extra reshuffling operation: As shown in Fig. \ref{fig:input3}, for each position in the 'after RB' memory, the data can 
come from 2 $(= PW_{b}/R^{cl}_{1,2})$ + 1 $(= PW_{b}/R^{cl}_{1,1})$ = 3 places, leading to 3 one-input MUXes in front of each word position of the 'after RB', therefore 12 MUXes are required in total. Or in general, approximately:
\begin{equation}
MUX_{buffer} \approx PW_b \cdot z \left( \displaystyle\sum \frac{PW_b}{min(PW_b, R^{cl}_{i,j})} \right) 
\end{equation}
where summation happens over unique values of $min(PW_{b}, R^{cl}_{i,j})$. 

\section*{Acknowledgements}
This work was supported in part by the European Research Council (ERC) under Agreement 101088865, in part by the European Union’s Horizon 2020 Program under Agreement 101070374,in part by the Flanders AI Research Program, and in part by the KU Leuven.

\begin{IEEEbiography}[{\includegraphics[width=1in,height=1.25in,clip,keepaspectratio]{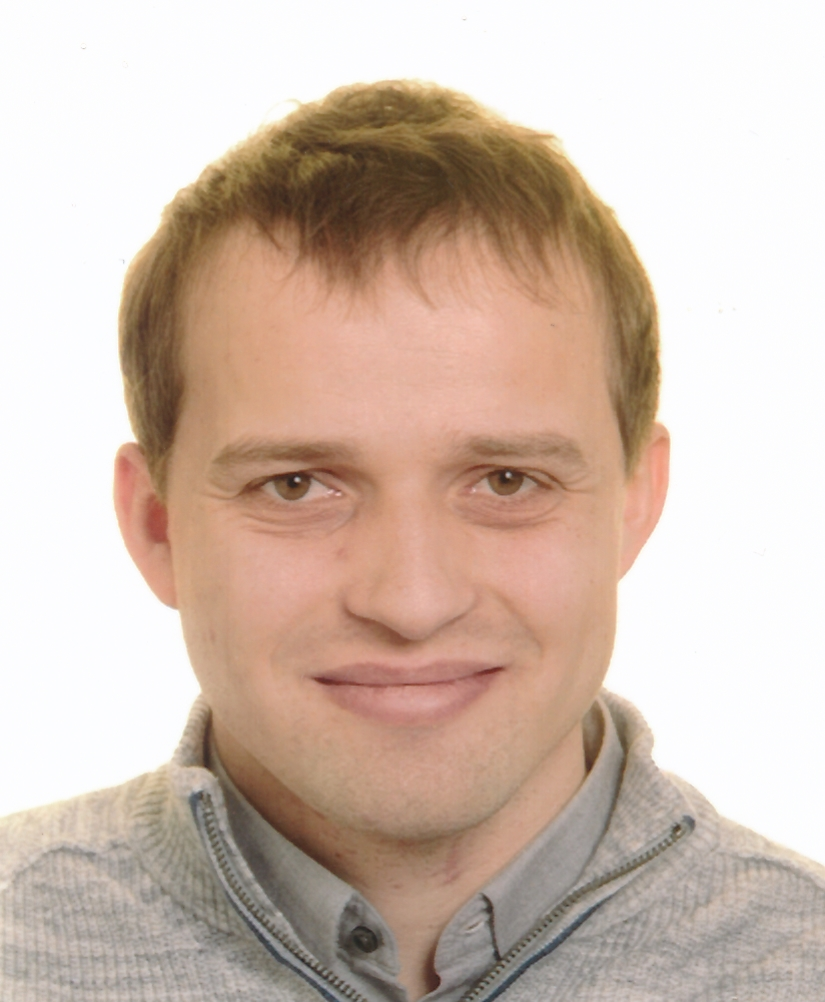}}]{Steven Colleman} Steven Colleman was born in Lier, Belgium in 1995. In 2018, he received the M.Sc. degree in electrical engineering from the KU Leuven, Belgium with the Master thesis: "Optimalisation of a Coarse Grain Reconfigurable Array for the efficient mapping of convolutional neural networks". Later in 2018, he started working as a PhD student at MICAS, in the lab of Prof. dr. ir. Marian Verhelst. His research interest lies in the field of efficient processing architectures for embedded deep learning. 
\end{IEEEbiography}

\begin{IEEEbiography}[{\includegraphics[width=1in,height=1.25in,clip,keepaspectratio]{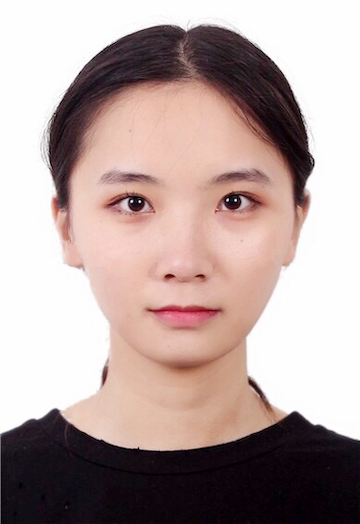}}]{Man Shi} Man Shi received the B.Sc. degree from the School of Information Science and Engineering from Shandong University (SDU), China, in 2017, and the M.Sc. degree from the Institute of Microelectronics, Tsinghua University, China, in 2020. She is currently pursuing the Ph.D. degree in the accelerators architecture for deep neural network with the ESAT-MICAS Laboratories, KU Leuven, Leuven, Belgium. Her current research interests include low-power deep neural network hardware accelerator design, algorithm-hardware co-design, and reconfigured computation.
\end{IEEEbiography}

\begin{IEEEbiography}[{\includegraphics[width=1in,height=1.25in,clip,keepaspectratio]{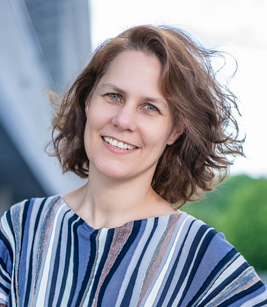}}]{Marian Verhelst} Marian Verhelst is a full professor at the MICAS laboratories of KU Leuven and a research director at imec. Her research focuses on embedded machine learning, hardware accelerators, HW-algorithm co-design and low-power edge processing. She received a PhD from KU Leuven in 2008, and worked as a research scientist at Intel Labs, Hillsboro OR from 2008 till 2010. Marian is a member of the board of directors of tinyML and active in the TPC’s of DATE, ISSCC, VLSI and ESSCIRC and was the chair of tinyML2021 and TPC co-chair of AICAS2020. Marian is an IEEE SSCS Distinguished Lecturer, was a member of the Young Academy of Belgium, an associate editor for TVLSI, TCAS-II and JSSC and a member of the STEM advisory committee to the Flemish Government. Marian received the laureate prize of the Royal Academy of Belgium in 2016, the 2021 Intel Outstanding Researcher Award, and the André Mischke YAE Prize for Science and Policy in 2021.
\end{IEEEbiography}

\end{document}